\documentclass[10pt,a4paper]{article}

\usepackage{jheppub}

\usepackage{amsmath}
\usepackage{amssymb}
\usepackage{graphicx,color}

\usepackage{amsmath,amsthm}
\usepackage{amssymb}

\usepackage{graphicx}
\usepackage{bm}

\newcommand{\Ref}[1]{(\ref{#1})}
\newtheorem{Theorem}{Theorem}[section]
\newtheorem{Definition}{Definition}[section]
\newtheorem{Lemma}[Theorem]{Lemma}%[section]

\newcommand{\startproof}{\textbf{Proof:\ \ }}
\newcommand{\finishproof}{\hfill $\Box$ \\}

\newcommand{\half}{\frac{1}{2}}
\newcommand{\Slc}{\mathrm{SL}(2,\mathbb{C})}
\newcommand{\slc}{\fs\fl_2\mathbb{C}}
\newcommand{\Uqsl}{\mathrm{U}_q(\fs\fl_2\mathbb{C})}

\newcommand{\tUqsl}{\tilde{\mathrm{U}}_q(\fs\fl_2\mathbb{C})}

\newcommand{\su}{\fs\fu_2}
\newcommand{\suq}{\mathrm{SU}_q(2)}
\newcommand{\Uqsu}{\mathrm{U}_q(\fs\fu_2)}

\def\be{\begin{eqnarray}}
\def\ee{\end{eqnarray}}

\newcommand{\ca}{\mathcal A}

\newcommand{\cb}{\mathcal B}
\newcommand{\cc}{\mathcal C}

\newcommand{\ck}{\mathcal K}

\newcommand{\cn}{\mathcal N}

\newcommand{\calr}{\mathcal R}

\newcommand{\fl}{\mathfrak{l}}

\newcommand{\fs}{\mathfrak{s}}  
  
\newcommand{\fu}{\mathfrak{u}}

\renewcommand{\a}{\alpha}
\renewcommand{\b}{\beta}
\newcommand{\g}{\gamma}
\newcommand{\G}{\Gamma}

\newcommand{\eps}{\varepsilon}

\newcommand{\sig}{\sigma}

\renewcommand{\l}{\lambda}
\renewcommand{\L }{\Lambda}
\renewcommand{\o}{\omega}

\renewcommand{\t}{\tau}

\newcommand{\rmd}{\mathrm d}

\newcommand{\lt}{\left}
\newcommand{\rt}{\right}

\newcommand{\lag}{\left\langle}
\newcommand{\rag}{\right\rangle}

\newcommand{\tr}{\mathrm{tr}}
\newcommand{\bbc}{\mathbb{C}}

\newcommand{\id}{\mathrm{id}}
\newcommand{\Pol}{\mathrm{Pol}}
\newcommand{\Fun}{\mathrm{Fun}}

\title{4-dimensional Spin-foam Model with Quantum Lorentz Group}

\author[]{Muxin Han}

\affiliation[]{Centre de Physique Th\'eorique%
\footnote{Unit\'e mixte de recherche (UMR 6207) du CNRS et des Universit\'es de Provence (Aix-Marseille I), de la Meditarran\'ee (Aix-Marseille II) et du Sud (Toulon-Var); laboratoire affili\'e \`a la FRUMAM (FR 2291).}, CNRS-Luminy Case 907,  F-13288 Marseille, France}

\emailAdd{Muxin.Han@cpt.univ-mrs.fr} %\emailAdd{buthor@univ.country} \emailAdd{cuthor@another.univ.country}

\abstract{We study the quantum group deformation of the Lorentzian EPRL spin-foam model. The construction uses the harmonic analysis on the quantum Lorentz group. We show that the quantum group spin-foam model so defined is free of the infra-red divergence, thus gives a finite partition function on a fixed triangulation. We expect this quantum group spin-foam model is a spin-foam quantization of discrete gravity with a cosmological constant.}

\begin{document}

\maketitle

\section{Introduction}

Loop Quantum Gravity (LQG) is an attempt to make a background independent, non-perturbative
quantization of 4-dimensional General Relativity (GR) -- for reviews, see \cite{book,rev,sfrevs}. It is
inspired by the classical formulation of GR as a dynamical theory of connections. Starting
from this formulation, the kinematics of LQG is well-studied and results in a successful kinematical
framework (see the corresponding chapters in the books \cite{book}), which is also unique in a
certain sense \cite{unique}. However, the framework of the dynamics in LQG is still largely open so
far. There are two main approaches to the dynamics of LQG, they are (1) the Operator formalism of
LQG, which follows the spirit of Dirac quantization or reduced phase space quantization of constrained dynamical system, and performs a canonical quantization of GR \cite{QSD}; (2) the Path integral formulation of LQG, which is currently understood in terms of the Spin-foam Models (SFMs) \cite{sfrevs,BC,EPRL,FK,HT}. The
relation between these two approaches is well-understood in the case of 3-dimensional gravity
\cite{perez}, while for 4-dimensional gravity, the situation is much more complicated and there are
some attempts \cite{links} for relating these two approaches.

The present article is concerning the framework of spin-foam models. The current spin-foam models for quantum gravity are mostly inspired by the 4-dimensional Plebanski formulation of GR \cite{plebanski} (Plebanski-Holst formulation by including the Barbero-Immirzi parameter $\g$), which is a BF theory constrained by the condition that the $B$ field should be ``simple'' i.e. there is a tetrad field $e^I$ such that $B={}^\star(e\wedge e)$. Currently one of the successful spin-foam model for Lorentzian signature is the EPRL spin-foam model defined in \cite{EPRL}, whose implementation of simplicity constraint is understood in the sense of \cite{DingYou}. The semiclassical limit of EPRL spin-foam model is shown to be well-behaved in the sense of \cite{semiclassical}.

However the EPRL model, as well as all the other spin-foam models defined with a classical group, suffer the problem of infra-red divergence, i.e. their full spin-foam amplitudes are always divergent because of the sum over face spins $j_f$, which takes values from zero to infinity. It is expected that this infra-red divergence will be regularized if we deform the spin-foam model from a classical group to a quantum group (see \cite{kassel} for an introduction of quantum group). For 3-dimensional gravity, the Turaev-Viro model \cite{TV} is a deformation of the Ponzano-Regge model \cite{PR} by the quantum group $\Uqsu$ ($q$ is a root of unity). The partition function of the Turaev-Viro model are finite and defines some invariants of 3-manifolds, because there is a quantum group cut-off of the admissible spin $j_f$ on each face. Moreover, the semiclassical limit of Turaev-Viro amplitude gives the 3-dimensional Regge action with a cosmological constant \cite{MT}. In 4-dimensions, the Crane-Yetter spin-foam model \cite{CY} is a deformation of 4-dimensional SU(2) BF theory (the Ooguri model \cite{Ooguri}) by $\Uqsu$ ($q$ is a root of unity). Similar to 3-dimensional case, the partition function of the Crane-Yetter model is finite and defines a topological invariant of 4-manifolds \cite{CKY}. Moreover the Crane-Yetter partition function is also the partition function of 4-dimensional SU(2) BF theory with a cosmological constant.

The lessons from 3-dimensional gravity and 4-dimensional topological field theory suggest that the quantum group deformation of 4-dimensional spin-foam models for quantum gravity will hopefully gives a finite partition function, which can be considered as a spin-foam model for quantum gravity with a cosmological constant. And it is also interesting to study the deformation of the Lorentzian spin-foam models by the non-compact quantum group $\Uqsl$ (quantum Lorentz group \cite{woro}). There are early pioneer works for q-deformed LQG \cite{smolin}. There are also an early investigation about the deformation of Barrett-Crane model by the quantum Lorentz group \cite{BCq}, where it is shown the deformation gave a finite spin-foam partition function.

In the present article, we study the deformation of the Lorentzian EPRL spin-foam model by the quantum Lorentz group $\Uqsl$. Finally we will show that thus a deformation gives a \emph{finite} spin-foam partition function. This work are motivated by the fact that the Lorentzian EPRL model has well-behaved semiclassical asymptotics in the sense of \cite{semiclassical}, thus there is a good chance for us to obtain a quantum group spin-foam model whose semiclassical limit recovers the discrete gravity with a cosmological constant (see the open problem No.16 in the first reference of \cite{sfrevs}).

Here is an outline of the article:

In Section 2 and Section 3, we review the facts about the compact quantum group $\Uqsu$ and the (non-compact) quantum Lorentz group $\Uqsl$. We also review the results about the harmonic analysis on the quantum Lorentz group \cite{roche}, which is the main mathematical tool for the present work.

In Section 4, we define a quantum group deformation of the Lorentzian EPRL intertwiner (quantum group relativistic intertwiner), which gives a well-defined $\Uqsl$ intertwiner.

In Section 5, we define a quantum group Lorentzian vertex amplitude, which is considered as a quantum group deformation of the Lorentzian EPRL vertex amplitude. And we show the finiteness of this q-Lorentzian vertex amplitude.

In Section 6, we write down a finite partition function for a spin-foam model with quantum Lorentz group.

\section{Compact Quantum Group $\Uqsu$}

First of all, we introduce some conventions and notations. we define a real deformation parameter $q=e^{-\o}\in\ ]0,1[$. The quantum groups recover the corresponding classical groups as $q\to1$ or $\o\to0$. Given an complex number $z\in\mathbb{C}$, the deformed q-number $[z]$ is defined by
\be
[z]=\frac{q^z-q^{-z}}{q-q^{-1}}
\ee
As $q\to1$, the deformed number recovers its classical limit, i.e. $\lim_{q\to1}[z]=z$. For any non-negative integer $n$, we can define the deformed factorials
\be
[0]!=1\ \ \ \ \ [n]!:=[1][2]\cdots[n]
\ee

We recall the definition of the quantum group $\Uqsu$ and review the basic facts about $\Uqsu$ (see \cite{kassel} for details. We follow the conventions and notations of \cite{roche})

%\begin{Definition}

The quantum group $\Uqsu$ is the universal enveloping algebra generated by four generators $q^{\pm J_z}$ and  $J_\pm$ with the algebraic relations
\be
q^{\pm J_z}q^{\mp J_z}=1\ \ \ \ \ q^{J_z}J_\pm q^{-J_z}=q^{\pm1}J_\pm\ \ \ \ \ \lt[J_+,\ J_-\rt]=\frac{q^{2J_z}-q^{-2J_z}}{q-q^{-1}}\label{qSU}
\ee
The comultiplication $\Delta:\ \Uqsu\to\Uqsu\otimes\Uqsu$, as an algebra morphism, is defined by
\be
\Delta(q^{\pm J_z})=q^{\pm J_z}\otimes q^{\pm J_z}\ \ \ \ \ \Delta(J_\pm)=q^{-J_z}\otimes J_\pm+J_\pm\otimes q^{J_z}
\ee
The counit $\eps:\ \Uqsu\to\bbc$, as an algebra morphism, is defined by
\be
\eps(J_\pm)=0\ \ \ \ \ \eps(q^{\pm J_z})=1
\ee
The antipode $S:\ \Uqsu\to\Uqsu^{op\ cop}$ \footnote{$H^{op}$ is the ``opposite'' Hopf algebra defined by reversing the multiplication of $H$, while $H^{cop}$ is the ``coopposite'' Hopf algebra defined by reversing the comultipliction of $H$.}, as a bialgebra morphism, is defined by
\be
S(J_+)=-J_+q^{-2J_z}\ \ \ \ S(J_-)=-q^{2J_z}J_-\ \ \ \ S(q^{\pm J_z})=q^{\mp J_z}
\ee
Moreover there is a $\star$-structure
\be
(q^{J_z})^\star=q^{J_z}\ \ \ \ \ J_{\pm}^\star=q^{\mp1}J_{\mp}
\ee
As $q\to1$, the quantum group $\Uqsu$ recovers the universal enveloping algebra $\text{U}(\su)$ \cite{kassel}.

%\end{Definition}

One can immediately check from the definition of the antipode that for all $x\in\Uqsu$
\be
S^2(x)=q^{2J_z}xq^{-2J_z}
\ee
The quantum group $\Uqsu$ is a ribbon quasi-triangular Hopf algebra. A quasi-triangular Hopf algebra $A$ has a extra structure which is a invertible element $R\in A\otimes A$ satisfying some conditions \cite{kassel}. $R$ is called a universal R-matrix. Let's write $R=\sum_i a_i\otimes b_i$, then the element $u=\sum_i S(b_i)a_i$ is invertible, its inverse $u^{-1}=\sum_i S^{-2}(b_i) a_i$ (assuming the antipode is invertible). Then we have for all $x\in A$
\be
S^2(x)=uxu^{-1}\ \ \ \ \ \Delta(u)=(u\otimes u)(R_{21}R)^{-1}=(R_{21}R)^{-1}(u\otimes u)
\ee
A quasi-triangular Hopf algebra $A$ is a ribbon Hopf $\star$-algebra, if and only if there exists an invertible central element $\theta$ such that
\be
\Delta(\theta)=(R_{21}R)^{-1}(\theta\otimes \theta)\ \ \ \ \eps(\theta)=1\ \ \ \ S(\theta)=\theta
\ee
A useful element $\mu$ is defined by $\mu=\theta^{-1}u$, it will be used to define the quantum trace, as will be seen later. For the quantum group $\Uqsu$, the universal R-matrix is given by\footnote{The R-matrix and the $\theta$ center are defined because $\Uqsu$ is defined as a topological Hopf algebra, see \cite{kassel} for precise discussion. }
\be
R=q^{2J_z\otimes J_z}e_{q^{-1}}^{(q-q^{-1})(q^{J_z}J_+\otimes J_-q^{-J_z})} \ \ \ \ \text{where}\ \ \ \ e_\a^z=\sum_{k=0}^\infty \a^{-\frac{k(k-1)}{2}}\frac{z^k}{[k]_\a!}
\ee
and the central element $\theta$ and the element $\mu$ is given by
\be
\theta=q^{C_q}\ \ \ \ \mu=q^{2J_z}
\ee
where $C_q$ is the \emph{quantum Casimir element} which will equal $-2K(K+1)$ on the unitary irreps of $\Uqsu$.

The unitary representations of $\Uqsu$ are completely reducible, and unitary irreps are completely classified by a couple $(\o,K)\in\{1.-1\}\times\frac{1}{2}\mathbb{N}$. The appearance of $\o$ comes from the existence of the automorphism $\t_\o$ of $\Uqsu$ defined by $\t_\o(q^{J_z})=\o q^{J_z}$, $\t_\o(J_+)=\o^2J_+$ and $\t_\o(J_-)=J_-$, which doesn't have classical counterpart. In the following we only consider the unitary irreps with $\o=1$. We denote by $Irr(\Uqsu)$ the set of all equivalent classes of unitary irreps with $\o=1$. Each of the unitary irreps $\pi^K\in Irr(\Uqsu)$ is labeled by a negative half-integer $K$ (a spin). The carrier space is denoted by $V_K$ which has dimension $2K+1$. The representation of $\Uqsu$ on $V_K$ is given concretely by
\be
\pi^K(q^{J_z})\ e^K_m=q^m\ e^K_m\ \ \ \ \ \pi^K(J_{\pm})\ e^K_m=q^{\mp1/2}\sqrt{[K\pm m+1][K\mp m]}\  e^K_m
\ee
where $e^K_m$ $(m=-K,\cdots,K)$ is the canonical basis. Given $M\in End(V_K)$, the \emph{quantum trace} $\tr_q(M)$ is defined by $\tr_q(M)=\tr_{V_j}(\mu^{-1} M)$. The definition of quantum trace is important for the proper definition of quantum group characters. In particular the q-dimension is given by $[d_K]=\tr_q(1)=[2K+1]$.

We have the Clebsch-Gordan decomposition of the tensor product representations, as the case of classical SU(2) group
\be
\pi^I\otimes\pi^J=\bigotimes_{K=|I-J|}^{I+J}\pi^K
\ee
For any three unitary irreps $\pi^I,\pi^J,\pi^K$, we define the Clebsch-Gordan maps $\psi^{K}_{IJ}\in \text{Hom}(V_I\otimes V_J,V_K)$ and $\phi_K^{IJ}\in\text{Hom}(V_K, V_I\otimes V_J)$. We define a function $Y(I,J,K)$
\be
Y(I,J,K)&=&1\ \ \ \ \text{if}\ \ \ \ I+J-K,\ J+K-I,\ K+I-J\in \mathbb{Z}^+\nonumber\\
Y(I,J,K)&=&0\ \ \ \ \text{otherwise}\label{Y}
\ee
When $Y(I,J,K)=0$, we have $\psi^K_{IJ}=\phi^{IJ}_K=0$. When $Y(I,J,K)\neq0$, $\psi^K_{IJ},\phi^{IJ}_K$ are nonzero and defined by the quantum Clebsch-Gordan coefficients:
\be
\phi^{IJ}_K(e^K_c)=\sum_{a,b}\left(
                               \begin{array}{cc}
                                 a & b  \\
                                 I & J \\
                               \end{array}
                             \Bigg|
                             \begin{array}{c}
                                 K   \\
                                 c \\
                               \end{array}
                             \right)e_a^I\otimes e^J_b\ \ \ \ \
\psi_{IJ}^K(e_a^I\otimes e^J_b)=\sum_c\left(\begin{array}{c}
                                 c   \\
                                 K \\
                               \end{array}
                             \Bigg|\begin{array}{cc}
                                 I & J  \\
                                 a & b \\
                               \end{array}\right)e^K_c
\ee
The detailed properties of the quantum Clebsch-Gordan coefficients is summarized in \cite{roche}.

We define the linear forms $u^K_{ab}=\lag e_a^K|\pi^K(\cdot)|e_b^K\rag$, from which we obtain a Hopf $\star$-algebra $\Pol(\Uqsu)$ of the polynomial functions on the quantum group $\Uqsu$:

%\begin{Definition}

The polynomial algebra $\mathrm{Pol}(\Uqsu)$ over $\Uqsu$ is a Hopf $\star$-algebra linearly spanned by $\{u^K_{ab}\}_{I\in\frac{\mathbb{N}}{2};a,b=-I\cdots I}$, with the following algebraic relations
\be
&&u^I_1u^J_2=\sum_K\phi^{IJ}_Ku^K\psi^K_{IJ}\ \ \ \ \text{which implies}\ \ \ \ R^{IJ}_{12}u_1^Iu_2^J=u_2^Ju_1^I R^{IJ}_{12}\nonumber\\
&&\Delta(u^K_{ab})=\sum_{c}u^K_{ac}\otimes u^K_{cb}\nonumber\\
&&\eps(u_{ab}^K)=\delta_{ab}\ \ \ \ \ \eta(1)=u^0\nonumber\\
&&S(u^K_{ab})=\sum_{c,d}w^K_{bc}\ u_{cd}^K\ w^{K\ -1}_{da}\ \ \ \ \text{where}\ \ \ \ w^K_{ab}=\delta_{a,-b}q^{a}(-1)^{K-a}\nonumber\\
&&(u^K_{ab})^\star=S(u^K_{ba})\ \ \ \ \text{i.e.}\ \ \ \ u^K (u^K)^\dagger=u^K (u^K)^\dagger=1
\ee
All the above algebraic relations realize that $\mathrm{Pol}(\Uqsu)$ is the dual Hopf $\star$-algebra of $\Uqsu$.
%\end{Definition}

$\Pol(\Uqsu)$ can equivalently be defined by the enveloping algebra of the matrix elements of $u^{1/2}$, with the above algebraic relations. More explicitly, $\Pol(\Uqsu)$ is the Hopf $\star$-algebra generated by the elements of
\be
u^{1/2}=\left(\begin{array}{cc}
a & b  \\
c & d \\
\end{array}\right)
\ee
satisfying the relations
\be
&&qab=ba\ \ \ \ qac=ca\ \ \ \ qbd=db\ \ \ \ qcd=dc\nonumber\\
&&bc=cb\ \ \ \ ad-da=(q^{-1}-q)bc\ \ \ \ ad-q^{-1}bc=1\nonumber\\
&&\Delta(a)=a\otimes a+b\otimes c\ \ \ \ \Delta(b)=b\otimes d+a\otimes b\ \ \ \ \Delta(c)=c\otimes a+d\otimes c\ \ \ \ \Delta(d)=d\otimes d+c\otimes b\nonumber\\
&&a^\star=d\ \ \ \ d^*=a\ \ \ \ b^\star=-q^{-1}c\ \ \ \ c^*=-qb\ \ \ \ S(u_{ab}^{1/2})=(u^{1/2}_{ba})^\star
\ee
In addition, we can define a norm on $\Pol(\Uqsu)$ by
\be
||x||:=\sup_{\pi}||\pi(x)||\ \ \ \ \ \forall\ x\in\Pol(\Uqsu)
\ee
where the supremum is taken over all unitary representations of $\Pol(\Uqsu)$. After the completion of $\Pol(\Uqsu)$ with this norm, we obtain a unital $C^\star$-algebra which contains the Hopf $\star$-algebra $\Pol(\Uqsu)$ as a dense domain. We denote the resulting $C^\star$-algebra by $\suq$, which is a example of compact matrix quantum group defined by Woronowicz \cite{compactquantumgroup}.

%\begin{Definition}
Given a Hopf $\star$-algebra $\ca$, a right invariant integral $h_R$ (or, left invariant integral $h_L$) of $\ca$ is a element in the dual Hopf algebra $\ca^*$, satisfying
\be
(h_R\otimes \id)\Delta=\eta\circ h_R\ \ \ \ &\text{or}&\ \ \ \ (\id\otimes h_L)\Delta=\eta\circ h_L\nonumber\\
h_R(a^\star)=\overline{h_R(a)}\ \ \ \ &\text{or}&\ \ \ \ h_L(a^\star)=\overline{h_L(a)}\ \ \ \ \ \  \forall\ a\in\ca
\ee
%\end{Definition}

For any compact quantum group, there exists a unique positive, normalized, left \emph{and} right invariant integral, namely a Haar integral \cite{compactquantumgroup}. For $\suq$, it is defined explicitly by
\be
h_{\suq}(u^K_{ab})=\delta_{K,0}.
\ee

\section{Quantum Lorentz Group}

Here we collect some basic facts about the quantum Lorentz group $\Uqsl$ with $q=e^{-\o}\in]0,1[$ and its harmonic analysis, see \cite{roche,woro} for details. We follow the same convention as \cite{roche}.

The quantum deformation of the $\slc$ enveloping algebra can be constructed via a \emph{quantum double} of $\Uqsu$, i.e.
\be
\Uqsl:=\Uqsu\ \hat{\otimes}\ \Pol(\Uqsu)^{cop}
\ee
where $\hat{\otimes}$ means that the $\Uqsu\otimes 1$ and $1\otimes \Pol(\Uqsu)^{cop}$ don't commutative. See e.g. \cite{kassel} for a detailed introduction for quantum double and the construction of universal R-matrix.

The unitary irreps of $\Uqsl$ is classified in \cite{pusz} (see also \cite{roche}). Here we only consider the principle series, which involves in the Plancherel theorem. The unitary irreps in the principle series are classified by a pair of numbers $\a=(m,\rho)$ where $m\in\mathbb{N} $ is a nonnegative integer and $\rho\in[-\frac{2\pi}{\o},\frac{2\pi}{\o}]\equiv I_\o$. We denote a principle series unitary irrep by $\stackrel{\a}{\Pi}$ and its carrier Hilbert space by $\stackrel{\a}{V}$. $\stackrel{\a}{V}$ is infinite dimensional and completely reducible with respect to the subalgebra generated by the $\Uqsu$ generators $J_\pm, q^{J_z}$, i.e.
\be
\stackrel{\a}{V}=\bigoplus_{K=\frac{k}{2}}^\infty\stackrel{I}{V}
\ee
where $\stackrel{I}{V}$ is the carrier space for the $\Uqsu$ unitary irrep labeled by the spin $I$.

One could define the space of the quantum analogue of functions on $\Uqsl$ ``vanishing at infinity''. Let's consider a dual space of $\Uqsl$, which is $\Uqsu^*\otimes\Pol(\Uqsu)^{*\ op}$. We define a dual basis $\{\stackrel{I}{E}\!{}^i_j\}$ in $\Pol(\Uqsu)^{*\ op}$ such that
\be
<\ \stackrel{I}{E}\!{}^i_j\ ,\ \stackrel{J}{u}\!{}^m_n\ >\ =\ \stackrel{I}{E}\!{}^i_j(\stackrel{J}{u}\!{}^m_n)\ =\ \delta_{IJ}\ \delta^i_n\ \delta^m_j
\ee
where we denote $\stackrel{J}{u}\!{}^i_j\equiv u^J_{ij}$. The multiplication between $\stackrel{K}{E}\!{}^i_j$ is dual to the comultiplication of $\stackrel{K}{u}\!{}^i_j$, so we can derive the following $\star$-algebra structure
\be
\stackrel{I}{E}\!{}^k_l \stackrel{J}{E}\!{}^r_s\ = \ \delta_{IJ} \stackrel{I}{E}\!{}^k_s \delta^r_l\ \ \ \ \ (\stackrel{I}{E}\!{}^i_j)^\star=\stackrel{I}{E}\!{}^j_i
\ee
which shows that the algebra generated by $\stackrel{J}{E}\!{}^i_j$ is isomorphic to the $\star$-algebra $\oplus_{I\in \mathbb{N}/2} \text{Mat}_{2I+1}(\bbc)$. It turns out that the elements in $\oplus_{I\in \mathbb{N}/2} \text{Mat}_{2I+1}(\bbc)$ are the quantum analogs of compact supported functions on the quantum hyperboloid $AN_q$, i.e
\be
\text{Fun}_c(AN_q)=\oplus_{I\in \mathbb{N}/2} \text{Mat}_{2I+1}(\bbc)
\ee
Note that this algebra has no unit element. One can endow this $\Fun_c(AN_q)$ a $C^\star$-norm by $||\oplus_Ia_I||:=\sup_I||a_I||$ where $||a_I||$ is the usual supremum norm for finite dimensional matrix. The $C^\star$ algebra $\Fun_0(AN_q)$, whose elements are the quantum analogs of functions vanishing at infinity, is defined by a completion of $\text{Fun}_c(AN_q)=\oplus_{I\in \mathbb{N}/2} \text{Mat}_{2I+1}(\bbc)$.

On the other hand, the comultiplication between $\stackrel{I}{E}\!{}^i_j$ can be computed by duality, which reads
\be
\Delta(\stackrel{I}{E}\!{}^i_j)=\sum_{J,K}\left(
                               \begin{array}{cc}
                                 n & s  \\
                                 J & K \\
                               \end{array}
                             \Bigg|
                             \begin{array}{c}
                                 I  \\
                                 j \\
                               \end{array}
                             \right)\left(\begin{array}{c}
                                 i   \\
                                 J \\
                               \end{array}
                             \Bigg|\begin{array}{cc}
                                 J & K  \\
                                 m & r \\
                               \end{array}\right)\stackrel{J}{E}\!{}^m_n\otimes\stackrel{K}{E}\!{}^r_s\label{comulti}
\ee
This comultiplication should be understood as the follows: $\Fun_c(AN_q)$ is endowed with a multiplier Hopf algebra \cite{multiplier}, whose comultiplication $\Delta$
\be
\Delta:\ \Fun_c(AN_q)\to M(\Fun_c(AN_q)\otimes \Fun_c(AN_q))
\ee
where $M(A)$ denote the algebra of multipliers of $A$. It is a consequence from the fact that we are considering a noncompact quantum group, $\Fun_c(AN_q)$ doesn't have a unit element. Therefore the infinite sum in Eq.(\ref{comulti}) should be understood in the sense of a multiplier.

up to this point we endow $\Pol(\Uqsu)$ a different $\star$-structure. We denote it by $\Pol(\Uqsu')$ the new $\star$-algebra, and denote the previous $\stackrel{J}{u}\!{}^i_j$ by $\stackrel{J}{k}\!{}^i_j$ as the basis of $\Pol(\Uqsu')$, while the new involution is defined by
\be
(\stackrel{J}{k}\!{}^i_j)^\star=S^{-1}(\stackrel{J}{k}\!{}^j_i)
\ee
Therefore we define the normed $\star$-algebra
\be
\Fun_{c}(\Uqsl):=\Pol(\Uqsu')\otimes\Fun_c(AN_q)
\ee
and the $C^\star$-algebra
\be
\Fun_{0}(\Uqsl):=\Fun(\Uqsu')\otimes\Fun_0(AN_q)
\ee
whose basis is denoted by
\be
\Big\{\stackrel{I}{k}\!{}^i_j\otimes \stackrel{J}{E}\!{}^k_l\Big\}_{I,J\in\frac{\mathbb{N}}{2};\ i,j\in\{-I\cdots I\};\ k,l\in\{-J\cdots J\}}
\ee
Moreover $\Fun_{c}(\Uqsl)$ is endowed with the structure of multiplier Hopf algebra, while its coalgebra structure on $\Pol(\Uqsu')\otimes\Fun_c(AN_q)$ has to be twisted, since it is dual to a quantum double. The detailed multiplier Hopf algebra structure of $\Fun_{c}(\Uqsl)$ is shown in \cite{roche,woro}. We denote $\Fun(\Uqsl)$ the set of affiliated elements to $\Fun_0(\Uqsl)$\footnote{Heuristically speaking, given a $C^\star$ algebra $A$, a linear operator $T:\ A\to A$ is a element affiliated to $A$, if the bounded functions of $T$ are bounded multipliers.}. We have
\be
\Fun(\Uqsl)=\prod_{I\in\frac{\mathbb{N}}{2}}\Fun(\Uqsu')\otimes\text{Mat}_{2I+1}(\bbc)
\ee
Here $\Fun(AN_q):=\prod_{I\in\frac{\mathbb{N}}{2}}\text{Mat}_{2I+1}(\bbc)$, whose elements are sequences $(a_I)_{I\in\frac{\mathbb{N}}{2}}$. They defines the left multipliers of $\Fun_0(AN_q)$ by the definition $(a_I)(\oplus_I b_I):=\oplus_Ia_Ib_I$. $\Fun(AN_q):=\prod_{I\in\frac{\mathbb{N}}{2}}\text{Mat}_{2I+1}(\bbc)$ is a Hopf $\star$-algebra by the relations
\be
&&(a_I)(b_I)=(a_Ib_I)\ \ \ \ \ \Delta(a_I)=\lt(\phi^{IJ}_Ia_I\psi^I_{JK}\rt)_{J,K}\nonumber\\
&&(a_I)^\star=(a_I^\dagger)\ \ \ \ \ \ \ \ \ \ \ \eta(1)=(1_I)_I
\ee
where $1_I$ is the $2I+1$ identity matrix.

$\Fun_c(AN_q)$ admits a right invariant integral defined by
\be
h_{AN_q}(\stackrel{I}{E}\!{}^i_j)=[d_I]\stackrel{I}{\mu}\!{}^{-1}{}^i_j
\ee
Then the left and right invariant Haar integral of $\Fun_c(\Uqsl)$ is given by $h=h_{\suq}\otimes h_{AN_q}$ and
\be
h(\stackrel{I}{k}\!{}^i_j\otimes \stackrel{J}{E}\!{}^k_l)=\delta_{I,0}[d_J]\stackrel{J}{\mu}\!{}^{-1}{}^k_l
\ee
With this Haar integral and the $\star$-structure on $\Fun_c(\Uqsl)$, we can define the $L^2$ inner product for any two elements in $\Fun_c(\Uqsl)$ and obtain a Hilbert space $L^2(\Uqsl)$ after completion.

The dual $\Fun_c(\Uqsl)^*$ is denoted by $\tUqsl$\footnote{$\tUqsl$ contains $\Uqsl$ as a Hopf subalgebra.}, which is also endowed with a multiplier Hopf algebra structure. Define the dual basis $x^\ca=\stackrel{I}{X}\!{}^i_j\otimes\stackrel{J}{g}\!{}^k_l$ in $\tUqsl$, dual to the basis $x_\ca=\stackrel{I}{k}\!{}^i_j\otimes\stackrel{J}{E}\!{}^k_l$ in $\Fun_c(\Uqsl)$, with the duality bracket
\be
<\stackrel{A_1}{X}\!{}^{a_1}_{a_1'}\otimes\stackrel{A_2}{g}\!{}^{a_2}_{a_2'},\ \stackrel{B_1}{k}\!{}^{b_1'}_{b_1}\otimes\stackrel{B_2}{E}\!{}^{b_2'}_{b_2}>\
:=\ \stackrel{A_1}{X}\!{}^{a_1}_{a_1'}\otimes\stackrel{A_2}{g}\!{}^{a_2}_{a_2'}\lt(\stackrel{B_1}{k}\!{}^{b_1'}_{b_1}\otimes\stackrel{B_2}{E}\!{}^{b_2'}_{b_2}\rt)
:=\delta^{A_1B_1}\delta^{A_2B_2}\delta^{a_1}_{b_1}\delta^{a_2}_{b_2}\delta_{a_1'}^{b_1}\delta_{a_2'}^{b_2'}
\ee

Given a principle unitary irrep $\stackrel{\a}{\Pi}$ labeled by $\a=(m,\rho)$, one can uniquely associate a unique representation $\stackrel{\a}{\tilde{\Pi}}$
\begin{eqnarray}
\stackrel{\alpha}{\tilde{\Pi}}(\stackrel{A}{X}\!{}^{a}_{a'}\otimes
\stackrel{B}{g}\!{}^{b}_{b'})\stackrel{C}{e}_{c}=
\stackrel{A}{e}_{a'}\sum_{D}\Lambda^{BD}_{AC}(\alpha)
\left(
                               \begin{array}{cc}
                                 a & b  \\
                                 A & B \\
                               \end{array}
                             \Bigg|
                             \begin{array}{c}
                                 D  \\
                                 d \\
                               \end{array}
                             \right)
\left(\begin{array}{c}
                                 d   \\
                                 D \\
                               \end{array}
                             \Bigg|\begin{array}{cc}
                                 B & C  \\
                                 b' & c \\
                               \end{array}\right)
\end{eqnarray}
where $\stackrel{C}{e}_{c}$ denotes the canonical basis for the $\Uqsu$ irreps. $\Lambda^{BD}_{AC}(\alpha)$ are coefficients defined in terms of analytic continuation of $6j$ symbols of $\Uqsu$ and whose properties are studied in depth in \cite{roche}. The representation matrix element of $\stackrel{\a}{\tilde{\Pi}}$ can be expanded in terms of the basis $\stackrel{I}{k}\!{}^i_j\otimes\stackrel{J}{E}\!{}^k_l$, i.e.
\be
\langle\ \stackrel{A}{e}_{a} |\ \stackrel{\alpha}{\tilde{\Pi}}(\cdot)\ | \stackrel{B}{e}_{b}\ \rangle=\sum_{C,D}\Lambda^{CD}_{AB}(\alpha)
\left(
                               \begin{array}{cc}
                                 a' & c'  \\
                                 A & C \\
                               \end{array}
                             \Bigg|
                             \begin{array}{c}
                                 D  \\
                                 d \\
                               \end{array}
                             \right)
\left(\begin{array}{c}
                                 d   \\
                                 D \\
                               \end{array}
                             \Bigg|\begin{array}{cc}
                                 C & B \\
                                 c & b \\
                               \end{array}\right)\stackrel{A}{k}\!{}^a_{a'}\otimes\stackrel{C}{E}\!{}^c_{c'}
\ee
Given $f\in\Fun_c(\Uqsl)$ (a smear function), we can define $\stackrel{\alpha}{\Pi}[f]$ to be a element of $End(\stackrel{\alpha}{V})$ by
\be
\stackrel{\alpha}{\Pi}[f]:=\sum_{AB}\stackrel{\alpha}{\tilde{\Pi}}\lt(\stackrel{A}{X}\!{}^{i}_{j}\otimes
\stackrel{B}{g}\!{}^{k}_{l}\rt)\ h\lt(\stackrel{A}{k}\!{}^{j}_{i}\otimes
\stackrel{B}{E}\!{}^{l}_{k}\cdot f\rt)
\ee
where the sum only involve a finite number of nonzero terms. And the matrix of $\stackrel{\alpha}{\Pi}[f]$ in the canonical basis $\stackrel{A}{e}_{a}$ ($A\in\frac{\mathbb{N}}{2}$, $a=-A,\cdots,A$) only has a finite number of nonzero matrix elements. As a result we can define the character by a quantum trace
\be
\chi_\a[f]=\tr_{\stackrel{\a}{V}}\lt(\stackrel{\alpha}{\Pi}(\mu^{-1})\stackrel{\alpha}{\Pi}[f]\rt)
\ee
which is shown to be well-defined \cite{roche} by its invariance properties.

The Plancherel formula for $\Uqsl$ is given by \cite{roche}
\be
\sum_{m=0}^\infty\int_{-\frac{2\pi}{\o}}^{\frac{2\pi}{\o}}P(m,\rho)\ \chi_{(m,\rho)}[f]\ \rmd\rho=\eps(f)
\ee
where $\eps$ is the counit of $\Fun_c(\Uqsl)$, and the Plancherel density $P(m,\rho)$ is given by
\be
P(m,\rho)=-\frac{\o}{2\pi}\Big[1-\frac{1}{2}\delta_{m,0}\Big]\Big[\cosh(m\o)-\cos(\o\rho)\Big].
\ee

In addition, the $\mathcal{R}$-matrix of $\Uqsl$ is given by the construction of quantum double
\be
\calr&=&\sum_A\stackrel{A}{X}\!{}^i_j\otimes 1\otimes 1\otimes\stackrel{A}{g}\!{}^j_i\nonumber\\
\calr^{-1}&=&\sum_A\stackrel{A}{X}\!{}^i_j\otimes 1\otimes 1\otimes\stackrel{A}{g}\!{}^j_i\circ S\label{R}
\ee
its matrix elements on principle unitary irreps are given by \cite{roche}
\be
\langle\ \stackrel{C}{e}_{c}\otimes \stackrel{D}{e}_{d} |\ \stackrel{\alpha}{\Pi}\otimes\stackrel{\b}{\Pi}(\mathcal{R})\ | \stackrel{A}{e}_{a}\otimes\stackrel{B}{e}_{b}\ \rangle&=&\delta_A^C\delta_a^c\sum_{M}\Lambda^{AM}_{DB}(\b)
\left(
                               \begin{array}{cc}
                                 d & c  \\
                                 D & C \\
                               \end{array}
                             \Bigg|
                             \begin{array}{c}
                                 M  \\
                                 m \\
                               \end{array}
                             \right)
\left(\begin{array}{c}
                                 m   \\
                                 M \\
                               \end{array}
                             \Bigg|\begin{array}{cc}
                                 A & B \\
                                 a & b \\
                               \end{array}\right)\nonumber\\
\langle\ \stackrel{C}{e}_{c}\otimes \stackrel{D}{e}_{d} |\ \stackrel{\alpha}{\Pi}\otimes\stackrel{\b}{\Pi}(\mathcal{R}^{-1})\ | \stackrel{A}{e}_{a}\otimes\stackrel{B}{e}_{b}\ \rangle
&=&\delta_A^C\delta_a^c\sum_{M}\Lambda^{AM}_{DB}(\b)
\left(
                               \begin{array}{cc}
                                 d & c'  \\
                                 D & C \\
                               \end{array}
                             \Bigg|
                             \begin{array}{c}
                                 M  \\
                                 m \\
                               \end{array}
                             \right)
\left(\begin{array}{c}
                                 m   \\
                                 M \\
                               \end{array}
                             \Bigg|\begin{array}{cc}
                                 A & B \\
                                 a' & b \\
                               \end{array}\right)\stackrel{A}{w}_{c'c}(\stackrel{A}{w}\!\!{}^{-1})^{aa'}\label{repR}
\ee
where $\stackrel{A}{w}_{c'c}=\delta_{c',-c}q^{-c}(-1)^{A-c'}$ and $(\stackrel{A}{w}\!\!{}^{-1})^{ca'}=\delta_{a',-c} q^c(-1)^{A-a'}$.

\section{Quantum Group Relativistic Intertwiner}

Now we generalize the definition of the Lorentzian EPRL intertwiner defined in \cite{EPRL} to the case of quantum Lorentz group. In \cite{BCq} the quantum group generalization has been done for the Barrett-Crane spin-foam model.

We denote the matrix element by a duality bracket
\be
\langle\ \stackrel{A}{e}_{a} |\ \stackrel{\alpha}{\Pi}(x)\ | \stackrel{B}{e}_{b}\ \rangle\ \equiv\ <\ \stackrel{\alpha}{\Pi}_{A,a;B,b}\big|\ x\ >
\ee
where $x\in \Uqsl$. The Lorentzian q-relativistic intertwiner is defined by a linear map from a $\Uqsu$ intertwiner to a $\Uqsl$ intertwiner:

\begin{Definition}
\begin{itemize}

\item Given a unitary irreps $A$ (nonnegative half-integers) of $\Uqsu$, we associate them with a principle unitary irreps $\a[A]$ of $\Uqsl$ by defining
\be
\a[A]=(m[A],\rho[A]):=(2A,2\g A)
\ee
where $\g$ is the Barbero-Immirzi parameter. Since $\rho\in[-\frac{2\pi}{\o},\frac{2\pi}{\o}]$, the spin $A$ has to be restricted in $A\in[-\frac{\pi}{|\g|\o},\frac{\pi}{|\g|\o}]$.

\item Given a $n$-valent $\Uqsu$ intertwiner $(n>2)$
\be
C[\mathbf{A}]\in \mathrm{Inv}_{\Uqsu}\big(\stackrel{A_1}{V}\otimes\cdots\otimes\stackrel{A_n}{V}\big)
\ee
with $n$ unitary irreps $\mathbf{A}=(A_1,\cdots,A_n)$ of $\Uqsu$, we define a Lorentzian q-relativistic intertwiner (or, a EPRL$_q$ intertwiner) $\l[\mathbf{A}]$ by
\be
\l[\mathbf{A}]_{B_1,b_1;\cdots;B_n,b_n}:=\sum_{\ca}C[\mathbf{A}]^{a_1\cdots a_n} <\ \bigotimes_{i=1}^n\stackrel{\a[A_i]}{\Pi\ }\!\!\!\!\!{}_{A_i,a_i;B_i,b_i}\ \big|\ \Delta^{(n)}x^{\ca}\ > h(x_\ca)\label{EPRLq}
\ee
where $x_\ca$ is a basis of $\Fun_c(\Uqsl)$ and $x^\ca$ is the dual basis. $h=h_{\Uqsu}\otimes h_{AN_q}$ is the Haar integral of $\Fun_c(\Uqsl)$.

\end{itemize}

\end{Definition}

The above definition is a quantum group deformation of the $n$-valent Lorentzian EPRL intertwiner \cite{EPRL} Note that in the above definition the expression
\be
&&\sum_\ca<\ \bigotimes_{i=1}^n\stackrel{\a_i}{\Pi }\!{}_{A_i,a_i;B_i,b_i}\ \big|\ \Delta^{(n)}x^{\ca}\ > h(x_\ca)\nonumber\\
&=&\sum_{\ca_1,\cdots,\ca_n}\prod_{i=1}^n<\ \stackrel{\a_i}{\Pi }\!{}_{A_i,a_i;B_i,b_i}\ \big|\ x^{\ca_i}\ > h(x_{\ca_1}\cdots x_{\ca_n})\label{integral}
\ee
is a quatum analog of the classical Lorentz group integral
\be
\int_{\Slc}\rmd g\ \prod_{i=1}^n\stackrel{\a_i}{\Pi }\!{}_{A_i,a_i;B_i,b_i}(g)
\ee
One can show the quantity Eq.(\ref{integral}) gives a invariant tensor under both left and right multiplication. We consider for example the right multiplication by $y\in\Uqsl$
\be
\sum_\ca<\ \bigotimes_{i=1}^n\stackrel{\a_i}{\Pi }\ \big|\ \Delta^{(n)}(x^{\ca}y)\ > h(x_\ca)
\ee
We define a new basis $\tilde{x}^\ca:=x^\ca y$ and denote its dual basis by $\tilde{x}_\ca$, then we have\footnote{Here we are using Sweedler's sigma notation.}
\be
<x^\ca,x_\cb>=\delta^\ca_\cb=<\tilde{x}^\ca,\tilde{x}_\cb>=<x^\ca\otimes y,\Delta x_\cb>=\sum_{(\tilde{x}_\cb)}<x^\ca,\tilde{x}^1_\cb><y,\tilde{x}^2_\cb>
\ee
therefore we obtain a transformation of dual basis $x_\cb=\sum_{(\tilde{x}_\cb)}\tilde{x}^1_\cb<y,\tilde{x}^2_\cb>$. By the right invariance of the Haar integral we have $h(x_\cb)=h(\tilde{x}_\cb)$. As a result
\be
\sum_\ca<\ \bigotimes_{i=1}^n\stackrel{\a_i}{\Pi }\ \big|\ \Delta^{(n)}(x^{\ca}y)\ > h(x_\ca)
&=&\sum_\ca<\ \bigotimes_{i=1}^n\stackrel{\a_i}{\Pi }\ \big|\ \Delta^{(n)}(\tilde{x}^{\ca})\ > h(\tilde{x}_\ca)\nonumber\\
&=&\sum_\ca<\ \bigotimes_{i=1}^n\stackrel{\a_i}{\Pi }\ \big|\ \Delta^{(n)}(x^{\ca})\ > h(x_\ca)
\ee
which shows the invariance under right multiplication. The invariance under left multiplication can be shown similarly.

Moreover we can show that the quantity Eq.(\ref{integral}) converges for $n>2$: We employ the basis $x_\ca=\stackrel{I}{k}\!{}^{j}_{i}\otimes\stackrel{J}{E}\!{}^{l}_{k}$ in $\Fun_c(\Uqsl)$ and its dual basis $x^\ca=\stackrel{I}{X}\!{}^{i}_{j}\otimes\stackrel{J}{g}\!{}^{k}_{l}$. Recall that
\be
h(\stackrel{I}{k}\!{}^j_i\otimes \stackrel{J}{E}\!{}^l_k)&=&\delta_{I,0}[d_J]\stackrel{J}{\mu}\!{}^{-1}{}^l_k\\
\langle\ \stackrel{A}{e}_{a} |\ \stackrel{\alpha}{{\Pi}}(\cdot)\ | \stackrel{B}{e}_{b}\ \rangle&=&\sum_{C,D}\Lambda^{CD}_{AB}(\alpha)
\left(
                               \begin{array}{cc}
                                 a' & c'  \\
                                 A & C \\
                               \end{array}
                             \Bigg|
                             \begin{array}{c}
                                 D  \\
                                 d \\
                               \end{array}
                             \right)
\left(\begin{array}{c}
                                 d   \\
                                 D \\
                               \end{array}
                             \Bigg|\begin{array}{cc}
                                 C & B \\
                                 c & b \\
                               \end{array}\right)\stackrel{A}{k}\!{}^a_{a'}\otimes\stackrel{C}{E}\!{}^c_{c'}
\ee
We can compute that
\be
\langle\ \stackrel{A}{e}_{a} |\ \stackrel{\alpha}{{\Pi}}(\stackrel{I}{X}\!{}^{i}_{j}\otimes\stackrel{J}{g}\!{}^{k}_{l})\ | \stackrel{B}{e}_{b}\rangle
&=&\sum_{C,D}\Lambda^{CD}_{AB}(\alpha)
\left(
                               \begin{array}{cc}
                                 a' & c'  \\
                                 A & C \\
                               \end{array}
                             \Bigg|
                             \begin{array}{c}
                                 D  \\
                                 d \\
                               \end{array}
                             \right)
\left(\begin{array}{c}
                                 d   \\
                                 D \\
                               \end{array}
                             \Bigg|\begin{array}{cc}
                                 C & B \\
                                 c & b \\
                               \end{array}\right)<\stackrel{A}{k}\!{}^a_{a'}\otimes\stackrel{C}{E}\!{}^c_{c'},\ \stackrel{I}{X}\!{}^{i}_{j}\otimes\stackrel{J}{g}\!{}^{k}_{l}>\nonumber\\
&=&\sum_{C,D}\Lambda^{CD}_{AB}(\alpha)
\left(
                               \begin{array}{cc}
                                 a' & c'  \\
                                 A & C \\
                               \end{array}
                             \Bigg|
                             \begin{array}{c}
                                 D  \\
                                 d \\
                               \end{array}
                             \right)
\left(\begin{array}{c}
                                 d   \\
                                 D \\
                               \end{array}
                             \Bigg|\begin{array}{cc}
                                 C & B \\
                                 c & b \\
                               \end{array}\right)\delta^{AI}\delta^a_j\delta^i_{a'}\delta^{CJ}\delta^k_{c'}\delta^c_l\nonumber\\
&=&\sum_{D}\Lambda^{JD}_{AB}(\alpha)
\left(
                               \begin{array}{cc}
                                 i & k  \\
                                 A & J \\
                               \end{array}
                             \Bigg|
                             \begin{array}{c}
                                 D  \\
                                 d \\
                               \end{array}
                             \right)
\left(\begin{array}{c}
                                 d   \\
                                 D \\
                               \end{array}
                             \Bigg|\begin{array}{cc}
                                 J & B \\
                                 l & b \\
                               \end{array}\right)\delta^{AI}\delta^a_j
\ee
On the other hand for the Haar integral:
\be
&&h\lt(\stackrel{I_1}{k}\!{}^{j_1}_{i_1}\otimes\stackrel{J_1}{E}\!{}^{l_1}_{k_1}\cdots\stackrel{I_n}{k}\!{}^{j_n}_{i_n}\otimes\stackrel{J_n}{E}\!{}^{l_n}_{k_n}\rt)\nonumber\\
&=&h_{\suq}\lt(\stackrel{I_1}{k}\!{}^{j_1}_{i_1}\cdots\stackrel{I_n}{k}\!{}^{j_n}_{i_n}\rt)\delta^{J_1J_2}\delta^{J_2J_3}\cdots\delta^{J_{n-1}J_n}\delta^{l_2}_{k_1}\delta^{l_3}_{k_2}\cdots\delta^{l_n}_{k_{n-1}}h_{AN_q}(\stackrel{J_n}{E}\!{}^{l_1}_{k_n})\nonumber\\
&=&h_{\suq}\lt(\stackrel{I_1}{k}\!{}^{j_1}_{i_1}\cdots\stackrel{I_n}{k}\!{}^{j_n}_{i_n}\rt)\delta^{J_1J_2}\delta^{J_2J_3}\cdots\delta^{J_{n-1}J_n}\delta^{l_2}_{k_1}\delta^{l_3}_{k_2}\cdots\delta^{l_n}_{k_{n-1}}[d_{J_n}]\stackrel{J_n}\mu\!\!{}^{-1}{}^{l_1}_{k_n}
\ee
Insert these result into the definition of the quantum group intertwiner $\l[\mathbf{A}]_{B_1,b_1;\cdots;B_n,b_n}$ we obtain its concrete expression
\be
&&\sum_{\ca_1\cdots\ca_n}C[\mathbf{A}]^{a_1\cdots a_n} \prod_{p=1}^n< \stackrel{\a[A_p]}{\Pi\ }\!\!\!\!\!{}_{A_p,a_p;B_p,b_p} \big|x^{\ca_p} > h(x_{\ca_1}\cdots x_{\ca_n})\nonumber\\
&=&\sum_{\{I_p,J_p,i_p,j_p,k_p,l_p\}}C[\mathbf{A}]^{a_1\cdots a_n} \prod_{p=1}^n\sum_{D_p}\Lambda^{J_pD_p}_{A_pB_p}(\alpha[A_p])
\left(
                               \begin{array}{cc}
                                 i_p & k_p  \\
                                 A_p & J_p \\
                               \end{array}
                             \Bigg|
                             \begin{array}{c}
                                 D_p  \\
                                 d_p \\
                               \end{array}
                             \right)
\left(\begin{array}{c}
                                 d_p   \\
                                 D_p \\
                               \end{array}
                             \Bigg|\begin{array}{cc}
                                 J_p & B_p \\
                                 l_p & b_p \\
                               \end{array}\right)\delta^{A_pI_p}\delta^{a_p}_{j_p}\nonumber\\
&&h_{\suq}\lt(\stackrel{I_1}{k}\!{}^{j_1}_{i_1}\cdots\stackrel{I_n}{k}\!{}^{j_n}_{i_n}\rt)\delta^{J_1J_2}\delta^{J_2J_3}\cdots\delta^{J_{n-1}J_n}\delta^{l_2}_{k_1}\delta^{l_3}_{k_2}\cdots\delta^{l_n}_{k_{n-1}}[d_{J_n}]\stackrel{J_n}\mu\!\!{}^{-1}{}^{l_1}_{k_n}\nonumber\\
&=&\sum_{\{i_p,l_p\},J,k_n}C[\mathbf{A}]^{a_1\cdots a_n} \prod_{p=1}^n\sum_{D_p}\Lambda^{JD_p}_{A_pB_p}(\alpha[A_p])
\left(
                               \begin{array}{cc}
                                 i_p & l_{p+1}  \\
                                 A_p & J \\
                               \end{array}
                             \Bigg|
                             \begin{array}{c}
                                 D_p  \\
                                 d_p \\
                               \end{array}
                             \right)
\left(\begin{array}{c}
                                 d_p   \\
                                 D_p \\
                               \end{array}
                             \Bigg|\begin{array}{cc}
                                 J & B_p \\
                                 l_p & b_p \\
                               \end{array}\right)\nonumber\\
&&h_{\suq}\lt(\stackrel{A_1}{k}\!{}^{a_1}_{i_1}\cdots\stackrel{A_n}{k}\!{}^{a_n}_{i_n}\rt)[d_{J}]\stackrel{J}\mu\!\!{}^{-1}{}^{l_1}_{k_n}
\ee
Since $C[\mathbf{A}]$ is a $\Uqsu$ intertwiner and by the normalization of $h_{\suq}$
\be
&&\l[\mathbf{A}]_{B_1,b_1;\cdots;B_n,b_n}\nonumber\\
&=&\sum_{J}\sum_{D_1\cdots D_n}\sum_{l_1\cdots l_n,l_{n+1}}\sum_{i_1\cdots i_n}C[\mathbf{A}]^{i_1\cdots i_n} \prod_{p=1}^n\Lambda^{JD_p}_{A_pB_p}(\alpha[A_p])
\left(
                               \begin{array}{cc}
                                 i_p & l_{p+1}  \\
                                 A_p & J \\
                               \end{array}
                             \Bigg|
                             \begin{array}{c}
                                 D_p  \\
                                 d_p \\
                               \end{array}
                             \right)
\left(\begin{array}{c}
                                 d_p   \\
                                 D_p \\
                               \end{array}
                             \Bigg|\begin{array}{cc}
                                 J & B_p \\
                                 l_p & b_p \\
                               \end{array}\right)[d_{J}]\stackrel{J}\mu\!\!{}^{-1}{}^{l_1}_{l_{n+1}}\label{1}
\ee
where we have relabeled $k_n\equiv l_{n+1}$. Note that there is only one infinite sum $\sum_J$ in the above formula, and all the other summations are within a finite range. We define
\be
\Theta\Big(\{A_p,i_p;B_p,b_p\},\{D_p\},J\Big)=\sum_{l_1\cdots l_n,l_{n+1}}\stackrel{J}\mu\!\!{}^{-1}{}^{l_1}_{l_{n+1}}\prod_{p=1}^n\left(
                               \begin{array}{cc}
                                 i_p & l_{p+1}  \\
                                 A_p & J \\
                               \end{array}
                             \Bigg|
                             \begin{array}{c}
                                 D_p  \\
                                 d_p \\
                               \end{array}
                             \right)
\left(\begin{array}{c}
                                 d_p   \\
                                 D_p \\
                               \end{array}
                             \Bigg|\begin{array}{cc}
                                 J & B_p \\
                                 l_p & b_p \\
                               \end{array}\right)
\ee
The Clebsch-Gordan coefficients are bounded by one, so we obtain the bound
\be
\lt|\Theta\Big(\{A_p,i_p;B_p,b_p\},\{D_p\},J\Big)\rt|\leqslant[d_J](2J+1)^{n-1}\prod_{p=1}^n(2D_p+1)
\ee
Moreover the asymptotics of $\Lambda^{BC}_{AD}(\alpha)$ has been studied in \cite{roche}. For a fixed $\a$, as $J$ goes to be large there exists $\cc_0(A_p,B_P,R_p)>0$
\be
|\Lambda^{J\ J+R_p}_{A_pB_p}(\alpha)| \leqslant \cc_0(A_p,B_p,R_p) Jq^{2J}
\ee
where $R_p\in\frac{1}{2}\mathbb{Z}$.

\begin{Lemma}\label{bound}

Given a $\Uqsl$ unitary irrep $\a=(m,\rho)$, and fix the index $B$ of $\L^{BC}_{AD}(\a)$, the absolute value of the coefficients $\L^{BC}_{AD}(\a)$, as a function of the labels $C,A,D$, is bounded by a linear function of $C$ as $C\neq 0$. It implies that the above $ \cc_0(A_p,B_p,R_p)$ is a linear function of $J+R_p$ and independent of $A_p,B_p$.

\end{Lemma}

\startproof We prove the lemma inductively. First of all we consider $\L^{BC}_{A}\equiv\L^{BC}_{AA}$. From \cite{roche} we know that
\be
\L^{BC}_{A}=\sum_{\sig_1,\sig_2}\left(\begin{array}{c}
                                 m   \\
                                 A \\
                               \end{array}
                             \Bigg|\begin{array}{cc}
                                 C & B \\
                                 \sig_2 & \sig_1 \\
                               \end{array}\right)\left(
                               \begin{array}{cc}
                                 \sig_1 & \sig_2 \\
                                 B & C \\
                               \end{array}
                             \Bigg|
                             \begin{array}{c}
                                 D  \\
                                 m \\
                               \end{array}
                             \right)q^{-2i\sig_1\rho}
\ee
Because the Clebsch-Gordan coefficients are bounded by 1, we have
\be
|\L^{BC}_{A}|\leq\sum_{\sig_1,\sig_2}1=(2C+1)(2B+1)
\ee
Therefore $|\L^{BC}_{A}|$ is bounded by a linear function of $C$ for a given $B$.

If we assume for a given $B$, $|\L^{BC}_{AD}|$ is bounded by a linear function of $C$, we consider $|\L^{BC}_{AD+1}|$. By using the following relation with quantum 6j-symbols \cite{roche}
\be
|\L^{B\ C}_{A\ D+1}\L^{\half\ D+\half}_{D+1\ D}|\leq\sum_{K,U}\frac{[d_U]^{\half}[d_{D+1}]^{\half}}{[d_C]^{\half}[d_{D+\half}]^{\half}}\lt|\left\{
                               \begin{array}{cc}
                                 \half & C \\
                                 A & D+\half \\
                               \end{array}
                             \Bigg|
                             \begin{array}{c}
                                 U  \\
                                 D+1 \\
                               \end{array}
                             \right\}\left\{
                               \begin{array}{cc}
                                 A & B \\
                                 \half & U \\
                               \end{array}
                             \Bigg|
                             \begin{array}{c}
                                 C  \\
                                 K \\
                               \end{array}
                             \right\}\left\{
                               \begin{array}{cc}
                                 B & \half \\
                                 D & U \\
                               \end{array}
                             \Bigg|
                             \begin{array}{c}
                                 K  \\
                                 D+\half \\
                               \end{array}
                             \right\}\rt||\L^{KU}_{AD}|
\ee
where the sum only contains four terms $U=C\pm\half$, $K=B\pm\half$. By the fact that the quantum 6j-symbols are bounded by 1, we obtain
\be
|\L^{B\ C}_{A\ D+1}\L^{\half\ D+\half}_{D+1\ D}|\leq\sum_{K,U}\frac{[d_U]^{\half}[d_{D+1}]^{\half}}{[d_C]^{\half}[d_{D+\half}]^{\half}}|\L^{KU}_{AD}|
\ee
$\frac{[d_{D+1}]^{\half}}{[d_{D+\half}]^{\half}}$ is bounded and $\frac{[d_U]^{\half}}{[d_C]^{\half}}$ is also bounded if $C\neq 0$. And from the explicit expression of $\L^{\half\ D+\half}_{D+1\ D}$ in \cite{roche} we can check that there are a upper bound and a lower bound $a_1,a_2>0$ such that
\be
a_2\leq |\L^{\half\ D+\half}_{D+1\ D}|\leq a_1
\ee
Therefore by the assumption that $|\L^{BC}_{AD}|$ is bounded by a linear function of $C$, we conclude that $|\L^{B\ C}_{A\ D+1}|$ is also bounded by a linear function of $C$.

\finishproof

Let's come back to Eq.\Ref{1} and the coefficients $\Lambda^{JD_p}_{A_pB_p}(\alpha[A_p])$. For each $D_p$, $D_p=\{|J-A_p|,\cdots,J+A_p\}$. Thus for each term in the sum $\sum_{D_p}$ we can use the about bound by setting $R_p=A_p,A_p-1,\cdots$.

As a consequence, there exists a quadratic function $\cc(D_p)>0$ and a integer $\cn\in\mathbb{Z}_+$, such that
\be
\Big|\l[\mathbf{A}]_{B_1,b_1;\cdots;B_n,b_n}\Big|&\leqslant&\sum_{J}\sum_{D_1\cdots D_n}\lt|[d_J]\sum_{i_1\cdots i_n}C[\mathbf{A}]^{i_1\cdots i_n}\Theta\Big(\{A_p,i_p;B_p,b_p\},\{D_p\},J\Big)\prod_{p=1}^n\Lambda^{JD_p}_{A_pB_p}(\alpha[A_p])\rt|\nonumber\\
&\leqslant&\sum_{J}\sum_{D_1\cdots D_n}\cc(D_p) J^{2n-1}q^{2J(n-2)}\ \leqslant\ \sum_{J}(\text{constant}) J^{\cn}q^{2J(n-2)}
\ee
which shows the series converges absolutely when $n>2$. So we see that $\l[\mathbf{A}]$ is a well-defined $\Uqsl$ intertwiner.

Given a Lorentzian quantum group relativistic intertwiner $\l[\mathbf{A}]$ depending on $n$ principle unitary irreps $(\a[A_1],\cdots,\a[A_n])$, in principle the intertwiner $\l[\mathbf{A}]$ depends on the ordering of the $n$ unitary irreps $(\a[A_1],\cdots,\a[A_n])$. More precisely, since we are dealing with a quasi-triangular Hopf algebra (actually even a ribbon Hopf algebra) with a universal $\mathcal{R}$-matrix, the $\Uqsl$-modules form a braided tensor category. Without losing generality, we consider an elementary braiding $\t=\t_{12}$ which is a flip of the first two irreps of $(\a[A_1],\cdots,\a[A_n])$. We then have a isomorphism of $\Uqsl$-modules $\stackrel{\a_1}{V}\otimes \stackrel{\a_2}{V}$ and $\stackrel{\a_2}{V}\otimes \stackrel{\a_1}{V}$:
\be
c_{\a_1,\a_2}:&& \stackrel{\a_1}{V}\otimes \stackrel{\a_2}{V}\to\stackrel{\a_2}{V}\otimes \stackrel{\a_1}{V}\nonumber\\
&& \stackrel{\a_1}{v_1}\otimes \stackrel{\a_2}{v_2}\mapsto c_{A_1,A_2}(\stackrel{\a_1}{v_1}\otimes \stackrel{\a_2}{v_2}):=\t_{12}{R}_{12}(\stackrel{\a_1}{v_1}\otimes \stackrel{\a_2}{v_2})
\ee
One can show that the isomorphism so defined is a solution of the Yang-Baxter equation \cite{kassel}. Under this braiding the interwiner transforms to
\be
&&\l[\t_{12}\mathbf{A}]\ c_{\a[A_1],\a[A_2]}\nonumber\\
&=&\sum_{\ca}C[\t_{12}\mathbf{A}]\ {\llcorner}\ <\ \bigotimes_{i=\t_{12}(1)}^{\t_{12}(n)}\stackrel{\a[A_i]}{\Pi\ }\!\!\!\!\!{}\ \ \big|\ \Delta^{(n)}x^{\ca}\ >\t_{12}\calr_{12}\ h(x_\ca)\nonumber\\
&=&\sum_{\ca}C[\t_{12}\mathbf{A}]\t_{12}\ {\llcorner}\ <\ \bigotimes_{i=1}^{n}\stackrel{\a[A_i]}{\Pi\ }\!\!\!\!\!{}\ \ \big|\ \t_{12}\Delta^{(n)}x^{\ca}\ >\calr_{12}\ h(x_\ca)\nonumber\\
&=&\sum_{\ca}C[\mathbf{A}]\ {\llcorner}\ \calr_{12}<\ \bigotimes_{i=1}^{n}\stackrel{\a[A_i]}{\Pi\ }\!\!\!\!\!{}\ \ \big|\ \Delta^{(n)}x^{\ca}\ >\ h(x_\ca)\label{braid}
\ee
where we have ignored the indices for the tensor contraction but inserted a symbol $\llcorner$ instead. And in the second step, we have used the flip relation $\t_{12}^{-1}(\stackrel{\a_2}{\Pi}\otimes\stackrel{\a_1}{\Pi})\t_{12}=\stackrel{\a_1}{\Pi}\otimes\stackrel{\a_2}{\Pi}$. In the third step, we have used the breading relation for $\calr$-matrix
\be
\t_{12}\circ\Delta(x)=\calr_{12}\Delta(x)\calr_{12}{}^{-1}
\ee
Therefore from Eq.(\ref{braid}) we see that the braiding of the intertwiner is determined by the action of the $\Uqsl$ universal $\calr$-matrix on the $\Uqsu$ intertwiner $C[\mathbf{A}]$. Restore the index notation:
\be
C[\mathbf{A}]\ {\llcorner}\ \calr_{12}=C[\mathbf{A}]^{a_1\cdots a_n}\big(\stackrel{\a[A_1]\a[A_2]}{\calr_{12}}\big)_{A_1,a_1;A_2,a_2}^{B_1,b_1;B_2,b_2}
\ee
Recall the representation of $\calr$-matrix Eq.(\ref{repR}), we see that for the nonzero matrix elements $\big(\stackrel{\a[A_1]\a[A_2]}{\calr_{12}}\big)_{A_1,a_1;A_2,a_2}^{B_1,b_1;B_2,b_2}$, $A_1=B_1$ and $a_1=b_1$, but $A_2,a_2$ in general can be different from $B_2,b_2$. So $C[\mathbf{A}]\ {\llcorner}\ \calr_{12}$ is in general \emph{not} a $\Uqsu$ intertwiner anymore. Therefore we conclude that the intertwiner $\l[\mathbf{A}]$ is not invariant under braiding, in contrast to the deformed Barrett-Crane intertwiner defined in \cite{BCq}.

\begin{figure}[h]
\begin{center}
\includegraphics[width=9cm]{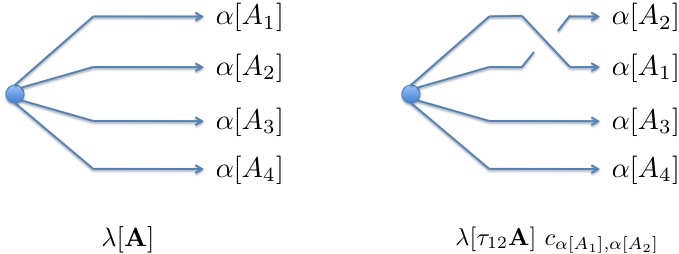}
\caption{A 4-valent q-relativistic intertwiner $\l[\mathbf{A}]$ and its braided intertwiner $\l[\t_{12}\mathbf{A}]\ c_{\a[A_1],\a[A_2]}$.}
\label{intertwiner}
\end{center}
\end{figure}

So far we are considering the $n$-valent intertwiners where all its legs are outgoing, as Fig.\ref{intertwiner}. However in general there is also intertwiners with both incoming and outgoing legs

\begin{Definition}

Given a $(p+n)$-valent $\Uqsu$ intertwiner $(n+p>2)$ with $p$ incoming legs and $n$ outgoing legs
\be
C[\mathbf{B},\mathbf{A}]\in \mathrm{Inv}_{\Uqsu}\big(\stackrel{B_1}{V}\!{}^\star\otimes\cdots\otimes\stackrel{B_p}{V}\!{}^\star\otimes\stackrel{A_1}{V}\otimes\cdots\otimes\stackrel{A_n}{V}\big)
\ee
where we have assumed a simplest ordering of the tensor product (in general one may have different ordering), we define the $(p+n)$-valent q-relativistic intertwiner $\l[\mathbf{B},\mathbf{A}]$ with $p$ incoming legs and $n$ outgoing legs by
\be
&&\l[\mathbf{B},\mathbf{A}]_{K_1,k_1\cdots K_p,k_p;J_1,j_1\cdots J_n,j_n}\nonumber\\
&:=&\sum_{\ca,\cb}< \bigotimes_{i=1}^p\stackrel{\a[B_i]}{\Pi\ }\!\!\!\!\!{}_{K_i,k_i;B_i,b_i} \big| \Delta^{op}{}^{(p)}S(x^{\cb}) >C[\mathbf{B},\mathbf{A}]^{b_1\cdots b_p,a_1\cdots a_n} < \bigotimes_{i=1}^n\stackrel{\a[A_i]}{\Pi\ }\!\!\!\!\!{}_{A_i,a_i;J_i,j_i} \big| \Delta^{(n)}x^{\ca} > h(x_\cb x_\ca)\label{EPRLq1}
\ee
where $x_\ca$ is a basis of $\Fun_c(\Uqsl)$ and $x^\ca$ is the dual basis, and $S$ is the antipode.

\end{Definition}

$\Delta^{op}{}^{(p)}$ is the comultiplication by reversing the ordering of $\Delta^{(p)}$. It is motivated by
\be
(S^{\otimes p}\otimes\id^{\otimes n})\circ\Delta^{(p+n)}=(\Delta^{op}{}^{(p)}\otimes\Delta^{(n)})\circ (S\otimes\id)\circ\Delta
\ee

It is straightforward to show that $\l[\mathbf{B},\mathbf{A}]$ is invariant under the right multiplication of $x^\ca$ by $y\in\Uqsl$. Indeed if we denote $\Delta y=\sum_y y^1\otimes y^2$ and if we make a change of basis $\tilde{x}^\cb=x^\cb y_1$ and $\tilde{x}^\ca=x^\ca y_2$, we have as before $x_\cb=\sum_{\tilde{x}_\cb}\tilde{x}^1_\cb<y_1,\tilde{x}^2_\cb>$ and $x_\ca=\sum_{\tilde{x}_\ca}\tilde{x}^1_\ca<y_2,\tilde{x}^2_\ca>$, then
\be
\sum_yx_\cb x_\ca=\sum_{\tilde{x}_\cb\tilde{x}_\ca}\tilde{x}^1_\cb\tilde{x}^1_\ca<y,\tilde{x}^2_\cb\tilde{x}^2_\ca>
\ee
As a result,
\be
&&\sum_{\ca,\cb}\sum_y< \bigotimes_{i=1}^p\stackrel{\a[B_i]}{\Pi\ }\!\!\!\!\!{}_{K_i,k_i;B_i,b_i} \big| \Delta^{op(p)}S(x^{\cb}y_1) >C[\mathbf{B},\mathbf{A}]^{b_1\cdots b_p,a_1\cdots a_n} < \bigotimes_{i=1}^n\stackrel{\a[A_i]}{\Pi\ }\!\!\!\!\!{}_{A_i,a_i;J_i,j_i} \big| \Delta^{(n)}(x^{\ca}y_2) > h(x_\cb x_\ca)\nonumber\\
&=&\sum_{\ca,\cb}< \bigotimes_{i=1}^p\stackrel{\a[B_i]}{\Pi\ }\!\!\!\!\!{}_{K_i,k_i;B_i,b_i} \big| \Delta^{op(p)}S(x^{\cb}) >C[\mathbf{B},\mathbf{A}]^{b_1\cdots b_p,a_1\cdots a_n} < \bigotimes_{i=1}^n\stackrel{\a[A_i]}{\Pi\ }\!\!\!\!\!{}_{A_i,a_i;J_i,j_i} \big| \Delta^{(n)}x^{\ca} > h(x_\cb x_\ca)
\ee
shows $\l[\mathbf{B},\mathbf{A}]$ is invariant under the right multiplication of $x^\ca$ by $y\in\Uqsl$.

By the properties of the duality bracket
\be
&&< \bigotimes_{i=1}^p\stackrel{\a[B_i]}{\Pi\ }\!\!\!\!\!{}_{K_i,k_i;B_i,b_i} \big| \Delta^{op(p)}S(x^{\cb}) >\nonumber\\
&=&< \stackrel{\a[B_p]}{\Pi\ }\!\!\!\!\!{}_{K_p,k_p;B_p,b_p}\stackrel{\a[B_{p-1}]}{\Pi\ }\!\!\!\!\!{}_{K_{p-1},k_{p-1};B_{p-1},b_{p-1}}\cdots\stackrel{\a[B_1]}{\Pi\ }\!\!\!\!\!{}_{K_1,k_1;B_1,b_1} \big|S(x^{\cb}) >\nonumber\\
&=&< S\big(\stackrel{\a[B_p]}{\Pi\ }\!\!\!\!\!{}_{K_p,k_p;B_p,b_p}\stackrel{\a[B_{p-1}]}{\Pi\ }\!\!\!\!\!{}_{K_{p-1},k_{p-1};B_{p-1},b_{p-1}}\cdots\stackrel{\a[B_1]}{\Pi\ }\!\!\!\!\!{}_{K_1,k_1;B_1,b_1} \big)\big|x^{\cb} >\nonumber\\
&=&<S\big(\stackrel{\a[B_1]}{\Pi\ }\!\!\!\!\!{}_{K_1,k_1;B_1,b_1}\big)\ S\big(\stackrel{\a[B_{2}]}{\Pi\ }\!\!\!\!\!{}_{K_{2},k_{2};B_{2},b_{2}}\big)\cdots S\big(\stackrel{\a[B_p]}{\Pi\ }\!\!\!\!\!{}_{K_p,k_p;B_p,b_p}\big) \big|x^{\cb} >
\ee
Because of the property of antipode on a matrix coalgebra $\sum_jS(\Pi_{ij})\Pi_{jk}=\delta_{ik}$ and the fact that $\stackrel{\a}{\Pi}$ is a unitary representation\footnote{A unitary representation $\pi$ of a Hopf $\star$-algebra satisfies $\pi(x)^\dagger=\pi(x^\star)$, then (in Sweedler's notation)
\be
\sum_x\pi_{ji}^\star(x_1) \pi_{jk}(x_2)=\sum_x\overline{\pi_{ji}(S^{-1}x^\star_1)} \pi_{jk}(x_2)=\sum_x\overline{\pi_{ji}(S(x_1)^\star)} \pi_{jk}(x_2)=\sum_x{\pi_{ij}(S(x_1))} \pi_{jk}(x_2)=\delta_{ik}\ \ \ \ \ (\text{summing over }j)
\ee
where we use the relations $\pi^\star(x)=\overline{\pi(S^{-1}x^\star)}$ and $S\circ\star=\star\circ S^{-1}$.}, we have ($\Lambda^{CD}_{BK}(\alpha)$ is real)
\be
S\big(\stackrel{\a}{\Pi}\!{}_{K,k;B,b}\big)&=&\stackrel{\a}{\Pi}\!{}_{B,b;K,k}^\star\nonumber\\
&=&\sum_{C,D}{\Lambda^{CD}_{BK}(\alpha)}
\left(
                               \begin{array}{cc}
                                 b' & c'  \\
                                 B & C \\
                               \end{array}
                             \Bigg|
                             \begin{array}{c}
                                 D  \\
                                 d \\
                               \end{array}
                             \right)
\left(\begin{array}{c}
                                 d   \\
                                 D \\
                               \end{array}
                             \Bigg|\begin{array}{cc}
                                 C & K \\
                                 c & k \\
                               \end{array}\right)(\stackrel{B}{k}\!{}^b_{b'}\otimes\stackrel{C}{E}\!{}^c_{c'})^\star\nonumber\\
&=&\sum_{C,D}{\Lambda^{CD}_{BK}(\alpha)}
\left(
                               \begin{array}{cc}
                                 b' & c'  \\
                                 B & C \\
                               \end{array}
                             \Bigg|
                             \begin{array}{c}
                                 D  \\
                                 d \\
                               \end{array}
                             \right)
\left(\begin{array}{c}
                                 d   \\
                                 D \\
                               \end{array}
                             \Bigg|\begin{array}{cc}
                                 C & K \\
                                 c & k \\
                               \end{array}\right)S^{-1}(\stackrel{B}{k}\!{}^{b'}_{b})\otimes\stackrel{C}{E}\!{}^{c'}_{c}
\ee
Then we can compute concretely:
\be
&&\l[\mathbf{B},\mathbf{A}]_{K_1,k_1\cdots K_p,k_p;J_1,j_1\cdots J_n,j_n}\nonumber\\
&=&\sum_{\{L_i,M_i,C_i,D_i\}}\prod_{i=1}^p{\Lambda^{L_iM_i}_{B_iK_i}(\alpha[B_i])}
\left(
                               \begin{array}{cc}
                                 b_i' & l_i'  \\
                                 B_i & L_i \\
                               \end{array}
                             \Bigg|
                             \begin{array}{c}
                                 M_i  \\
                                 m_i \\
                               \end{array}
                             \right)
\left(\begin{array}{c}
                                 m_i   \\
                                 M_i \\
                               \end{array}
                             \Bigg|\begin{array}{cc}
                                 L_i & K_i \\
                                 l_i & k_i \\
                               \end{array}\right)\prod_{i=1}^n\Lambda^{C_iD_i}_{A_iJ_i}(\alpha[A_i])
\left(
                               \begin{array}{cc}
                                 a_i' & c_i'  \\
                                 A_i & C_i \\
                               \end{array}
                             \Bigg|
                             \begin{array}{c}
                                 D_i  \\
                                 d_i \\
                               \end{array}
                             \right)
\left(\begin{array}{c}
                                 d_i   \\
                                 D_i \\
                               \end{array}
                             \Bigg|\begin{array}{cc}
                                 C_i & J_i \\
                                 c_i & j_i \\
                               \end{array}\right)\nonumber\\
&&C[\mathbf{B},\mathbf{A}]^{b_1\cdots b_p,a_1\cdots a_n}\ h_{\suq}(\prod_{i=1}^pS^{-1}(\stackrel{B_i}{k}\!{}^{b_i'}_{b_i})\prod_{i=1}^n\stackrel{A_i}{k}\!{}^{a_i}_{a_i'})\ h_{AN_q}(\prod_{i=1}^p\stackrel{L_i}{E}\!{}^{l_i'}_{l_i}\prod_{i=1}^n\stackrel{C_i}{E}\!{}^{c_i}_{c_i'})\nonumber\\
&=&\sum_{\{L_i,M_i,C_i,D_i\}}\prod_{i=1}^p{\Lambda^{L_iM_i}_{B_iK_i}(\alpha[B_i])}
\left(
                               \begin{array}{cc}
                                 b_i' & l_i'  \\
                                 B_i & L_i \\
                               \end{array}
                             \Bigg|
                             \begin{array}{c}
                                 M_i  \\
                                 m_i \\
                               \end{array}
                             \right)
\left(\begin{array}{c}
                                 m_i   \\
                                 M_i \\
                               \end{array}
                             \Bigg|\begin{array}{cc}
                                 L_i & K_i \\
                                 l_i & k_i \\
                               \end{array}\right)\prod_{i=1}^n\Lambda^{C_iD_i}_{A_iJ_i}(\alpha[A_i])
\left(
                               \begin{array}{cc}
                                 a_i' & c_i'  \\
                                 A_i & C_i \\
                               \end{array}
                             \Bigg|
                             \begin{array}{c}
                                 D_i  \\
                                 d_i \\
                               \end{array}
                             \right)
\left(\begin{array}{c}
                                 d_i   \\
                                 D_i \\
                               \end{array}
                             \Bigg|\begin{array}{cc}
                                 C_i & J_i \\
                                 c_i & j_i \\
                               \end{array}\right)\nonumber\\
&&C[\mathbf{B},\mathbf{A}]^{b'_1\cdots b'_p,a'_1\cdots a'_n}\ \delta^{L_1L_2}\cdots\delta^{L_{p-1}L_p}\delta^{C_1C_2}\cdots\delta^{C_{n-1}C_n}\delta^{l_2'}_{l_1}\cdots\delta^{l_p'}_{l_{p-1}}\delta^{c_2}_{c_1'}\cdots\delta^{c_n}_{c_{n-1}'}\delta^{L_pC_n}\delta^{c_1}_{l_p}\ h_{AN_q}(\stackrel{C_n}{E}\!{}^{l_1'}_{c_n'})\nonumber\\
&=&\sum_{C,\{M_i,D_i\}}\prod_{i=1}^p{\Lambda^{CM_i}_{B_iK_i}(\alpha[B_i])}
\left(
                               \begin{array}{cc}
                                 b_i' & l_{i-1}  \\
                                 B_i & C \\
                               \end{array}
                             \Bigg|
                             \begin{array}{c}
                                 M_i  \\
                                 m_i \\
                               \end{array}
                             \right)
\left(\begin{array}{c}
                                 m_i   \\
                                 M_i \\
                               \end{array}
                             \Bigg|\begin{array}{cc}
                                 C & K_i \\
                                 l_i & k_i \\
                               \end{array}\right)\prod_{i=1}^n\Lambda^{CD_i}_{A_iJ_i}(\alpha[A_i])
\left(
                               \begin{array}{cc}
                                 a_i' & c_{i+1}  \\
                                 A_i & C \\
                               \end{array}
                             \Bigg|
                             \begin{array}{c}
                                 D_i  \\
                                 d_i \\
                               \end{array}
                             \right)
\left(\begin{array}{c}
                                 d_i   \\
                                 D_i \\
                               \end{array}
                             \Bigg|\begin{array}{cc}
                                 C & J_i \\
                                 c_i & j_i \\
                               \end{array}\right)\nonumber\\
&&C[\mathbf{B},\mathbf{A}]^{b'_1\cdots b'_p,a'_1\cdots a'_n}\ [d_C]\stackrel{C}{\mu}\!{}^{-1}{}^{l_0}_{c_{n+1}}\ \delta^{c_1}_{l_p}
\ee
where in the second step we use the fact that $C[\mathbf{B},\mathbf{A}]$ is a $\Uqsu$ invariant tensor, and the $\suq$ Haar integral is normalized. Similar to the previous analysis, we can show that the above infinite sum converges absolutely, and $\l[\mathbf{B},\mathbf{A}]$ is not invariant under braiding. To summarize, we collect the results in this subsection as a theorem:

\begin{Theorem}

The $(p+n)$-valent q-relativistic intertwiner $\l[\mathbf{B},\mathbf{A}]$ is a well-defined (finite) $\Uqsl$ intertwiner. It is a invariant tensor under right multiplication of $\Uqsl$. And it is not invariant under braiding.

\end{Theorem}

\section{A Finite q-Lorentzian Vertex Amplitudes}

Given a boundary spin-network graph $\g$, roughly speaking, a q-Lorentzian vertex amplitude $A_v$ is defined by the following procedure: we associate each node $n$ of the graph a q-relativistic intertwiner $\l_n[\mathbf{B},\mathbf{A}]$, where each outgoing/incoming leg of the intertwiner is associated with a outgoing/incomming link connecting with the node. According to the way how different nodes are linked to one another, we contract the q-relativistic intertwiners to one another, while for each crossing we should use the isomorphism $c_{\a_1,\a_2}:\ \stackrel{\a_1}{V}\otimes \stackrel{\a_2}{V}\to\stackrel{\a_2}{V}\otimes \stackrel{\a_1}{V}$ to reverse the order of the irreps. More precisely:

\begin{itemize}

\item We assign a preferred direction, say from the left to the right of this paper. Then we draw the spin-network graph $\g$ on the paper by ordering the nodes from left to right and connect the nodes by linkes.

\item We associate each link $l$ of the graph $\g$ a $\Uqsl$ unitary irrep $\a_l$, $l\in L(\g)$ (the set of links of $\g$). For each link $l$ oriented from left to right, we make the following operation of the representation matrix element, and associate the link $l$ with the following function:
\be
(\id\otimes S)\Delta\stackrel{\a_l}{\Pi}_{A,a;B;b}\ =\ \sum_{C,c}\stackrel{\a_l}{\Pi}_{A,a;C;c}\otimes S\big(\stackrel{\a_l}{\Pi}_{C,c;B;b}\big)\ \equiv\ \stackrel{\a_l}{\Pi}_{(1)}\otimes S\big(\stackrel{\a_l}{\Pi}_{(2)}\big)
\ee
where we have employed the Sweedler's sigma notation for the comultiplcation, and ignore the symbol for sum. Therefore the two factors $\stackrel{\a_l}{\Pi}_{(1)}$ and $S\big(\stackrel{\a_l}{\Pi}_{(2)}\big)$ natually associated with the half-links, if we break the link $l$ into two half-links.

\item Suppose there is a crossing between two links $l_1$ and $l_2$, and $l_1$ and $l_2$ don't share their end-points, we associate the following function to these two links
\be
&&\Big[\stackrel{\a_1}{\Pi}_{(1)}\otimes S\big(\stackrel{\a_1}{\Pi}_{(3)}\big)\Big]\otimes\Big[\stackrel{\a_2}{\Pi}_{(1)}\otimes S\big(\stackrel{\a_2}{\Pi}_{(3)}\big)\Big]<\stackrel{\a_1}{\Pi}_{(2)}\otimes\stackrel{\a_2}{\Pi}_{(2)},\calr_{12}>\nonumber\\
\text{or}\ \ \
&&\Big[\stackrel{\a_1}{\Pi}_{(1)}\otimes S\big(\stackrel{\a_1}{\Pi}_{(3)}\big)\Big]\otimes\Big[\stackrel{\a_2}{\Pi}_{(1)}\otimes S\big(\stackrel{\a_2}{\Pi}_{(3)}\big)\Big]<\stackrel{\a_2}{\Pi}_{(2)}\otimes\stackrel{\a_1}{\Pi}_{(2)},\calr_{12}^{-1}>
\ee
When $l_1$ and $l_2$ share their end-point, the situation is the same as it was described previously (see Fig.\ref{intertwiner})

\item Whenever there is a link oriented from right to left, we associate it with the element $\mu^{-1}=q^{-2J_z}$, then the function associated to this link is
\be
S\big(\stackrel{\a}{\Pi}_{(3)}\big)\otimes \stackrel{\a}{\Pi}_{(1)}\ <\stackrel{\a}{\Pi}_{(2)},\mu^{-1}>
& =& \stackrel{\a}{\Pi}\lt(x^\ca\mu^{-1}S(x^\cb)\rt)x_{\cb}\otimes x_{\ca}\ =\ \stackrel{\a}{\Pi}\lt(\mu^{-1}S^2(x^\ca)S(x^\cb)\rt)x_{\cb}\otimes x_{\ca}\nonumber\\
&=&<\stackrel{\a}{\Pi}_{(1)},\mu^{-1}>S(\stackrel{\a}{\Pi}_{(2)})(x^\cb S(x^\ca))x_{\cb}\otimes x_{\ca}\nonumber\\
&=&<\stackrel{\a}{\Pi}_{(1)},\mu^{-1}>\big[S(\stackrel{\a}{\Pi}_{(2)})\big]_{(1)}\otimes S\Big(\big[S(\stackrel{\a}{\Pi}_{(2)})\big]_{(2)}\Big)\label{mu0}
\ee

\item Collect the functions for all the links, make a (ordered) multiplication over the half-links sharing the same begin or final point, and make tensor produces over all the different half-links without sharing end-point. In the end we have $|N(\g)|$ tensor product factors, where $|N(\g)|$ denotes the number of nodes of the graph $\g$. The resulting function is called a relativistic q-spin-network $f_{\g}$.

\item The vertex amplitude is defined by taking the $(|N(\g)|-1)$-fold Haar integral of the relativistic q-spin-network, i.e.
\be
A_v[\vec{K},\vec{\nu}]:=\big(\id\otimes h^{\otimes(|N(\g)|-1)}\big)\big(f_\g\big){\lrcorner}\Big(C_{\nu_1}\otimes \cdots\otimes C_{\nu_{|N(\g)|}}\Big)
\ee
where $C_\nu$ is a basis vector in the space of $\Uqsu$ intertwiners, and the $\Uqsu$ q-spin-network label $[\vec{K},\vec{\nu}]$ shows that the vertex amplitude so defined is a function of $\Uqsu$ q-spin-networks. Note that in contrast to the BC$_q$ vertex amplitude defined in \cite{BCq}, $A_v[\vec{K},\vec{\nu}]$ depends explicitly on the braidings of relativistic q-spin-network $f_{\g}$. The vertex amplitude $A_v$ can be also called a ``spin-foam \emph{quantum} trace'' over $N(\g)$ q-relativistic intertwiners, here the word \emph{quantum} refers to the appearance of the element $\mu^{-1}$ in the contraction.

\end{itemize}

\begin{figure}[h]
\begin{center}
\includegraphics[width=11cm]{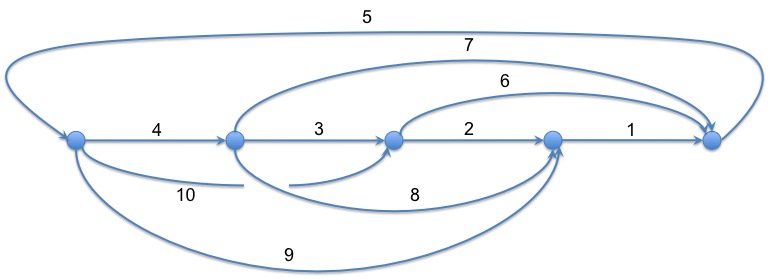}
\caption{The $\G_5^+$ graph in \cite{BCq}}
\label{gamma5}
\end{center}
\end{figure}

For example, we consider the $\G^+_5$ in Fig.\ref{gamma5} and write down its relativistic q-spin-network
\be
f_{\G_5^+}&:=&\Big[\big[S(\stackrel{\a_5}{\Pi}_{(2)})\big]_{(1)}\stackrel{\a_4}{\Pi}_{(1)}\stackrel{\a_{10}}{\Pi}_{(1)}\stackrel{\a_{9}}{\Pi}_{(1)}\Big]
\otimes \Big[S\big(\stackrel{\a_4}{\Pi}_{(2)}\big)\stackrel{\a_7}{\Pi}_{(1)}\stackrel{\a_3}{\Pi}_{(1)}\stackrel{\a_{8}}{\Pi}_{(1)}\Big]
\otimes\Big[S\big(\stackrel{\a_{10}}{\Pi}_{(3)}\big)S\big(\stackrel{\a_3}{\Pi}_{(2)}\big)\stackrel{\a_6}{\Pi}_{(1)}\stackrel{\a_{2}}{\Pi}_{(1)}\Big]\nonumber\\
&&\otimes\Big[S\big(\stackrel{\a_9}{\Pi}_{(2)}\big)S\big(\stackrel{\a_8}{\Pi}_{(3)}\big)S\big(\stackrel{\a_2}{\Pi}_{(2)}\big)\stackrel{\a_{1}}{\Pi}_{(1)}\Big]
\otimes\Big[S\big(\stackrel{\a_{1}}{\Pi}_{(2)}\big)S\big(\stackrel{\a_6}{\Pi}_{(2)}\big)S\big(\stackrel{\a_7}{\Pi}_{(2)}\big)S\Big(\big[S(\stackrel{\a_5}{\Pi}_{(2)})\big]_{(2)}\Big)\Big]<\stackrel{\a_{5}}{\Pi}_{(1)},\mu^{-1}>
\nonumber\\
&&<\stackrel{\a_{10}}{\Pi}_{(2)}\otimes\stackrel{\a_{8}}{\Pi}_{(2)},\calr> \label{app1}
\ee
then by the invariance of Haar integral and the braiding relation $\stackrel{\a_1\a_2}{\calr}_{12}\stackrel{\a_{1}}{\Pi}_1\stackrel{\a_{2}}{\Pi}_2=\stackrel{\a_{2}}{\Pi}_2\stackrel{\a_{1}}{\Pi}_1\stackrel{\a_1\a_2}{\calr}_{12}$ (see appendix for details)
\be
&&(\id\otimes h^{\otimes4})f_{\G_5^+}\nonumber\\
&=&h\Big[S\big(\stackrel{\a_4}{\Pi}\big)\stackrel{\a_7}{\Pi}_{(1)}\stackrel{\a_3}{\Pi}_{(1)}\stackrel{\a_{8}}{\Pi}_{(1)}\Big]
h\Big[S\big(\stackrel{\a_{10}}{\Pi}_{(2)}\big)S\big(\stackrel{\a_3}{\Pi}_{(2)}\big)\stackrel{\a_6}{\Pi}_{(1)}\stackrel{\a_{2}}{\Pi}_{(1)}\Big]\nonumber\\
&&h\Big[S\big(\stackrel{\a_9}{\Pi}\big)S\big(\stackrel{\a_8}{\Pi}_{(3)}\big)S\big(\stackrel{\a_2}{\Pi}_{(2)}\big)\stackrel{\a_{1}}{\Pi}_{(1)}\Big]
h\Big[S\big(\stackrel{\a_{5}}{\Pi}_{(2)}\big)\ \stackrel{\a_7}{\Pi}_{(2)}\stackrel{\a_6}{\Pi}_{(2)}\stackrel{\a_{1}}{\Pi}_{(2)}\Big]<\stackrel{\a_{5}}{\Pi}_{(1)},\mu^{-1}>
<\stackrel{\a_{10}}{\Pi}_{(1)}\otimes\stackrel{\a_{8}}{\Pi}_{(2)},\calr>\label{app2}
\ee
From the previous computation we have immediately (we ignore the factors which can be absorbed in to $\Uqsu$ intertwiners and doesn't contribute the infinite sum)
\be
&&{}^{B_4A_7A_3A_8}_{b_4\ a_7\ a_3\ a_8}h\Big[S\big(\stackrel{\a_4}{\Pi}\big)\stackrel{\a_7}{\Pi}_{(1)}\stackrel{\a_3}{\Pi}_{(1)}\stackrel{\a_{8}}{\Pi}_{(1)}\Big]{}^{A_4J_7J_3J_8}_{a_4\ j_7\ j_3\ j_8}\nonumber\\
&=&\sum_{C,\{M_i,D_i\}}{\Lambda^{CM_4}_{B_4A_4}(\alpha_4)}
\left(
                               \begin{array}{cc}
                                 b_4 & l  \\
                                 B_4 & C \\
                               \end{array}
                             \Bigg|
                             \begin{array}{c}
                                 M_4  \\
                                 m_4 \\
                               \end{array}
                             \right)
\left(\begin{array}{c}
                                 m_4   \\
                                 M_4 \\
                               \end{array}
                             \Bigg|\begin{array}{cc}
                                 C & A_4 \\
                                 l_4 & a_4 \\
                               \end{array}\right)\Lambda^{CD_7}_{A_7J_7}(\alpha_7)
\left(
                               \begin{array}{cc}
                                 a_7 & c_3  \\
                                 A_7 & C \\
                               \end{array}
                             \Bigg|
                             \begin{array}{c}
                                 D_7  \\
                                 d_7 \\
                               \end{array}
                             \right)
\left(\begin{array}{c}
                                 d_7   \\
                                 D_7 \\
                               \end{array}
                             \Bigg|\begin{array}{cc}
                                 C & J_7 \\
                                 l_4 & j_7 \\
                               \end{array}\right)\nonumber\\
&&\Lambda^{CD_3}_{A_3J_3}(\alpha_3)
\left(
                               \begin{array}{cc}
                                 a_3 & c_8  \\
                                 A_3 & C \\
                               \end{array}
                             \Bigg|
                             \begin{array}{c}
                                 D_3  \\
                                 d_3 \\
                               \end{array}
                             \right)
\left(\begin{array}{c}
                                 d_3   \\
                                 D_3 \\
                               \end{array}
                             \Bigg|\begin{array}{cc}
                                 C & J_3 \\
                                 c_3 & j_3 \\
                               \end{array}\right)\Lambda^{CD_8}_{A_8J_8}(\alpha_8)
\left(
                               \begin{array}{cc}
                                 a_8 & c  \\
                                 A_8 & C \\
                               \end{array}
                             \Bigg|
                             \begin{array}{c}
                                 D_8  \\
                                 d_8 \\
                               \end{array}
                             \right)
\left(\begin{array}{c}
                                 d_8   \\
                                 D_8 \\
                               \end{array}
                             \Bigg|\begin{array}{cc}
                                 C & J_8 \\
                                 c_8 & j_8 \\
                               \end{array}\right)\ [d_C]\stackrel{C}{\mu}\!{}^{-1}{}^{l}_{c}\nonumber\\
&\leqslant&\sum_{C,\{M_i,D_i\}}{\Lambda^{CM_4}_{B_4A_4}(\alpha_4)}
\Lambda^{CD_7}_{A_7J_7}(\alpha_7)
\Lambda^{CD_3}_{A_3J_3}(\alpha_3)
\Lambda^{CD_8}_{A_8J_8}(\alpha_8)
\ [d_C]^2\ \times \nonumber\\
&&\ \ \ \ \ \ \ \ \times(2C+1)^3\ (2M_4+1)\ (2D_7+1)\ (2D_3+1)\ (2D_8+1)
\ee

\be
&&{}^{K_{10}K_3A_6A_2}_{k_{10}\ k_3\ a_6\ a_2}h\Big[S\big(\stackrel{\a_{10}}{\Pi}_{(2)}\big)S\big(\stackrel{\a_3}{\Pi}_{(2)}\big)\stackrel{\a_6}{\Pi}_{(1)}\stackrel{\a_{2}}{\Pi}_{(1)}\Big]{}^{B_{10}B_3J_3J_2}_{b_{10}\ b_3\ j_6\ j_2}\nonumber\\
&=&\sum_{C,\{M_i,D_i\}}{\Lambda^{CM_{10}}_{B_{10}K_{10}}(\alpha_{10})}
\left(
                               \begin{array}{cc}
                                 b_{10}' & l  \\
                                 B_{10} & C \\
                               \end{array}
                             \Bigg|
                             \begin{array}{c}
                                 M_{10}  \\
                                 m_{10} \\
                               \end{array}
                             \right)
\left(\begin{array}{c}
                                 m_{10}   \\
                                 M_{10} \\
                               \end{array}
                             \Bigg|\begin{array}{cc}
                                 C & K_{10} \\
                                 l_{10} & k_{10} \\
                               \end{array}\right)
                               {\Lambda^{CM_{3}}_{B_{3}K_{3}}(\alpha_{3})}
\left(
                               \begin{array}{cc}
                                 b_{3}' & l_{10}  \\
                                 B_{3} & C \\
                               \end{array}
                             \Bigg|
                             \begin{array}{c}
                                 M_{3}  \\
                                 m_{3} \\
                               \end{array}
                             \right)
\left(\begin{array}{c}
                                 m_{3}   \\
                                 M_{3} \\
                               \end{array}
                             \Bigg|\begin{array}{cc}
                                 C & K_{3} \\
                                 l_{3} & k_{3} \\
                               \end{array}\right)\nonumber\\
&&\Lambda^{CD_6}_{A_6J_6}(\alpha_6)
\left(
                               \begin{array}{cc}
                                 a_6' & c_{2}  \\
                                 A_6 & C \\
                               \end{array}
                             \Bigg|
                             \begin{array}{c}
                                 D_6  \\
                                 d_6 \\
                               \end{array}
                             \right)
\left(\begin{array}{c}
                                 d_6   \\
                                 D_6 \\
                               \end{array}
                             \Bigg|\begin{array}{cc}
                                 C & J_6 \\
                                 l_3 & j_6 \\
                               \end{array}\right)
\Lambda^{CD_2}_{A_2J_2}(\alpha_2)
\left(
                               \begin{array}{cc}
                                 a_2' & c  \\
                                 A_2 & C \\
                               \end{array}
                             \Bigg|
                             \begin{array}{c}
                                 D_2  \\
                                 d_2 \\
                               \end{array}
                             \right)
\left(\begin{array}{c}
                                 d_2   \\
                                 D_2 \\
                               \end{array}
                             \Bigg|\begin{array}{cc}
                                 C & J_2 \\
                                 c_2 & j_2 \\
                               \end{array}\right)\ [d_C]\stackrel{C}{\mu}\!{}^{-1}{}^{l}_{c}\nonumber\\
&\leqslant&\sum_{C,\{M_i,D_i\}}{\Lambda^{CM_{10}}_{B_{10}K_{10}}(\alpha_{10})}
{\Lambda^{CM_{3}}_{B_{3}K_{3}}(\alpha_{3})}
\Lambda^{CD_6}_{A_6J_6}(\alpha_6)
\Lambda^{CD_2}_{A_2J_2}(\alpha_2)
\ [d_C]^2\ \times \nonumber\\
&&\ \ \ \ \ \ \ \ \times(2C+1)^3\ (2M_{10}+1)\ (2M_3+1)\ (2D_6+1)\ (2D_2+1)
\ee

\be
&&{}^{K_{9}K_8K_2A_1}_{k_{9}\ k_8\ k_2\ a_1}h\Big[S\big(\stackrel{\a_9}{\Pi}\big)S\big(\stackrel{\a_8}{\Pi}_{(3)}\big)S\big(\stackrel{\a_2}{\Pi}_{(2)}\big)\stackrel{\a_{1}}{\Pi}_{(1)}\Big]{}^{B_{9}B_8B_2J_1}_{b_{9}\ b_8\ b_2\ j_1}\nonumber\\
&=&\sum_{C,\{M_i,D_i\}}{\Lambda^{CM_9}_{B_9K_9}(\alpha_9)}
\left(
                               \begin{array}{cc}
                                 b_9' & l  \\
                                 B_9 & C \\
                               \end{array}
                             \Bigg|
                             \begin{array}{c}
                                 M_9  \\
                                 m_9 \\
                               \end{array}
                             \right)
\left(\begin{array}{c}
                                 m_9   \\
                                 M_9 \\
                               \end{array}
                             \Bigg|\begin{array}{cc}
                                 C & K_9 \\
                                 l_9 & k_9 \\
                               \end{array}\right){\Lambda^{CM_8}_{B_8K_8}(\alpha_8)}
\left(
                               \begin{array}{cc}
                                 b_8' & l_9  \\
                                 B_8 & C \\
                               \end{array}
                             \Bigg|
                             \begin{array}{c}
                                 M_8  \\
                                 m_8 \\
                               \end{array}
                             \right)
\left(\begin{array}{c}
                                 m_8   \\
                                 M_8 \\
                               \end{array}
                             \Bigg|\begin{array}{cc}
                                 C & K_8 \\
                                 l_8 & k_8 \\
                               \end{array}\right)\nonumber\\
&&{\Lambda^{CM_2}_{B_2K_2}(\alpha_2)}
\left(
                               \begin{array}{cc}
                                 b_2' & l_8  \\
                                 B_2 & C \\
                               \end{array}
                             \Bigg|
                             \begin{array}{c}
                                 M_2  \\
                                 m_2 \\
                               \end{array}
                             \right)
\left(\begin{array}{c}
                                 m_2   \\
                                 M_2 \\
                               \end{array}
                             \Bigg|\begin{array}{cc}
                                 C & K_2 \\
                                 l_2 & k_2 \\
                               \end{array}\right)\Lambda^{CD_1}_{A_1J_1}(\alpha_1)
\left(
                               \begin{array}{cc}
                                 a_1' & c  \\
                                 A_1 & C \\
                               \end{array}
                             \Bigg|
                             \begin{array}{c}
                                 D_1  \\
                                 d_1 \\
                               \end{array}
                             \right)
\left(\begin{array}{c}
                                 d_1   \\
                                 D_1 \\
                               \end{array}
                             \Bigg|\begin{array}{cc}
                                 C & J_1 \\
                                 l_2 & j_1 \\
                               \end{array}\right)\ [d_C]\stackrel{C}{\mu}\!{}^{-1}{}^{l}_{c}\nonumber\\
&\leqslant&\sum_{C,\{M_i,D_i\}}{\Lambda^{CM_9}_{B_9K_9}(\alpha_9)}
{\Lambda^{CM_8}_{B_8K_8}(\alpha_8)}
{\Lambda^{CM_2}_{B_2K_2}(\alpha_2)}
\Lambda^{CD_7}_{A_1J_1}(\alpha_1)
\ [d_C]^2\ \times \nonumber\\
&&\ \ \ \ \ \ \ \ \times(2C+1)^3\ (2M_9+1)\ (2M_8+1)\ (2M_2+1)\ (2D_1+1)
\ee

\be
&&{}^{K_{5}A_7A_6A_1}_{k_{5}\ a_7\ a_6\ a_1}h\Big[S\big(\stackrel{\a_{5}}{\Pi}_{(2)}\big) \stackrel{\a_7}{\Pi}_{(2)}\stackrel{\a_6}{\Pi}_{(2)}\stackrel{\a_{1}}{\Pi}_{(2)}\Big]{}^{B_{5}J_7J_6J_1}_{b_{5}\ j_7\ j_6\ j_1}\nonumber\\
&=&\sum_{C,\{M_i,D_i\}}\Lambda^{CM_5}_{K_5B_5}(\alpha_5)
\left(
                               \begin{array}{cc}
                                 k_5' & l  \\
                                 K_5 & C \\
                               \end{array}
                             \Bigg|
                             \begin{array}{c}
                                 M_5  \\
                                 m_5 \\
                               \end{array}
                             \right)
\left(\begin{array}{c}
                                 m_5   \\
                                 M_5 \\
                               \end{array}
                             \Bigg|\begin{array}{cc}
                                 C & B_5 \\
                                 l_5 & b_5 \\
                               \end{array}\right)\Lambda^{CD_7}_{A_7J_7}(\alpha_7)
\left(
                               \begin{array}{cc}
                                 a_7' & c_{6}  \\
                                 A_7 & C \\
                               \end{array}
                             \Bigg|
                             \begin{array}{c}
                                 D_7  \\
                                 d_7 \\
                               \end{array}
                             \right)
\left(\begin{array}{c}
                                 d_7   \\
                                 D_7 \\
                               \end{array}
                             \Bigg|\begin{array}{cc}
                                 C & J_7 \\
                                 l_5 & j_7 \\
                               \end{array}\right)\nonumber\\
&&\Lambda^{CD_6}_{A_6J_6}(\alpha_6)
\left(
                               \begin{array}{cc}
                                 a_6' & c_{1}  \\
                                 A_6 & C \\
                               \end{array}
                             \Bigg|
                             \begin{array}{c}
                                 D_6  \\
                                 d_6 \\
                               \end{array}
                             \right)
\left(\begin{array}{c}
                                 d_6   \\
                                 D_6 \\
                               \end{array}
                             \Bigg|\begin{array}{cc}
                                 C & J_6 \\
                                 c_6 & j_6 \\
                               \end{array}\right)\Lambda^{CD_1}_{A_1J_1}(\alpha_1)
\left(
                               \begin{array}{cc}
                                 a_1' & c  \\
                                 A_1 & C \\
                               \end{array}
                             \Bigg|
                             \begin{array}{c}
                                 D_1  \\
                                 d_1 \\
                               \end{array}
                             \right)
\left(\begin{array}{c}
                                 d_1   \\
                                 D_1 \\
                               \end{array}
                             \Bigg|\begin{array}{cc}
                                 C & J_1 \\
                                 c_1 & j_1 \\
                               \end{array}\right)\ [d_C]\stackrel{C}{\mu}\!{}^{-1}{}^{l}_{c}\nonumber\\
&\leqslant&\sum_{C,\{M_i,D_i\}}\Lambda^{CM_5}_{K_5B_5}(\alpha_5)
\Lambda^{CD_7}_{A_7J_7}(\alpha_7)
\Lambda^{CD_6}_{A_6J_6}(\alpha_6)
\Lambda^{CD_1}_{A_1J_1}(\alpha_1)
\ [d_C]^2\times \nonumber\\
&&\ \ \ \ \ \ \ \ \times(2C+1)^3\ (2M_5+1)\ (2D_7+1)\ (2D_6+1)\ (2D_1+1)
\ee

\be
{}^{A_{10}J_2}_{a_{10}j_2}<\stackrel{\a_{10}}{\Pi}_{(1)}\otimes\stackrel{\a_{8}}{\Pi}_{(2)},\calr>{}^{J_{10}K_2}_{j_{10}k_2}
&=&\delta^{A_{10}}_{J_{10}}\delta_{j_{10}}^{a_{10}}\sum_{M}\Lambda^{A_{10}M}_{J_2K_2}(\a_8)
\left(
                               \begin{array}{cc}
                                 j_2 & a_{10}  \\
                                 J_2 & A_{10} \\
                               \end{array}
                             \Bigg|
                             \begin{array}{c}
                                 M  \\
                                 m \\
                               \end{array}
                             \right)
\left(\begin{array}{c}
                                 m   \\
                                 M \\
                               \end{array}
                             \Bigg|\begin{array}{cc}
                                 A_{10} & K_2 \\
                                 j_{10} & k_2 \\
                               \end{array}\right)\nonumber\\
&\leqslant&\delta^{A_{10}}_{J_{10}}\delta_{j_{10}}^{a_{10}}\sum_{M}\Lambda^{A_{10}M}_{J_2K_2}(\a_8)(2M+1)\label{RR}
\ee
\be
{}^{A_5}_{a_5}<\stackrel{\a_{5}}{\Pi}_{(1)},\mu^{-1}>{}^{J_5}_{j_5}=\delta^{A_5J_5}\delta_{a_5j_5}q^{-2a_5}\label{mu1}
\ee
The above $\leqslant$ means the bounds of the absolute values, where we use the fact that the Clebsch-Gordan coefficients are bounded by 1. The integral $(\id\otimes h^{\otimes4})f_{\G_5^+}$ is given by the product of the above factors and sum over $\{M_i,D_i\}$, $J_i$, $K_i$ and four $C$'s (some $J_i$ and $K_i$ are contracted). We can observe the following results:

\begin{itemize}

\item As $C$ goes to be large there exists a linear function $\cc_0(C+R)>0$
\be
|\Lambda^{C\ C+R}_{AB}(\alpha)| \leqslant \cc_0(C+R)\ Cq^{2C}=\cc_0(C+R)\ Ce^{-2\o C}
\ee
for any $R\in\frac{1}{2}\mathbb{Z}$ $(C+R_i=D_i)$ \cite{roche}. Since for each $D_i$, $D_i=\{|C-A_i|,\cdots,C+A_i\}$, thus for each term in the sum $\sum_{D_i}$ we can use the about bound by setting $R_i=A_i,A_i-1,\cdots$. The same argument applies to the sum of $M_i$.

\item When $C$ goes to be large $[d_C]^2\sim q^{-4C}=e^{4\o C}$.

\item Each $M_i$ satisfies $|B_i-C|\leq M_i\leq B_i+C$, and each $D_i$ satisfies $|A_i-C|\leq D_i\leq A_i+C$, so they are bounded linearly by $C$'s.
\end{itemize}
From these results we know that, as $C$ goes to be large, each of the above integration contributes a
\be
\sum_C (\text{constant})C^{k}e^{-4\o C}
\ee
for some power $k>0$. In addition, we also can see that

\begin{itemize}
\item The factors
\be
\left(
                               \begin{array}{cc}
                                 a' & c  \\
                                 A & C \\
                               \end{array}
                             \Bigg|
                             \begin{array}{c}
                                 D  \\
                                 d \\
                               \end{array}
                             \right)
\left(\begin{array}{c}
                                 d   \\
                                 D \\
                               \end{array}
                             \Bigg|\begin{array}{cc}
                                 C & J \\
                                 c & j \\
                               \end{array}\right)
\ee
are nonzero only when there is a overlap between $\{A+C,\cdots,|A-C|\}$ and $\{C+J,\cdots,|C-J|\}$, so when $J$ goes to be large it has to be $|J-C|\leq A+C$, which gives $J\leq2C+A$. Thus each $J_i$ is linearly bounded by $C$. And the same argument and result applies to each $K_i$.

\item About the representation of the $\calr$-matrix Eq.(\ref{RR}), from Lemma \ref{bound}, $|\Lambda^{A_{10}M}_{J_2K_2}(\a_8)|$ is bounded by a linear function $\cc(M)$ while $\cc(M)$ is independent of $J_2,K_2$. Since $M$ lies in the intersection of $\{|J_2-A_{10}|,\cdots, J_2+A_{10}\}$ and $\{|K_2-A_{10}|,\cdots, K_2+A_{10}\}$, the sum $|\sum_M\Lambda^{A_{10}M}_{J_2K_2}(\a_8)(2M+1)|$ is bounded by a third order polynomial function of $J_2$ or $K_2$

\item The sum over $j_i$ or $k_i$ gives $(2J_i+1)$ or $(2K_i+1)$.

\end{itemize}

Therefore we conclude that there exists integers $\cn_1,\cn_2,\cn_3,\cn_4$
\be
|(\id\otimes h^{\otimes4})f_{\G_5^+}|\leqslant\sum_{C_1,C_2,C_3,C_4}(\text{constant})\ e^{-4\o C_1}e^{-4\o C_2}e^{-4\o C_3}e^{-4\o C_4}\ C^{\cn_1}_1C_2^{\cn_2}C_3^{\cn_3}C_4^{\cn_4}
\ee
which converges absolutely. To summarize:

\begin{Theorem}
The $\G^+_5$ relativistic q-spin-network is integrable, and the vertex amplitude defined by $\G^+_5$ relativistic q-spin-network is finite.
\end{Theorem}

\begin{figure}[h]
\begin{center}
\includegraphics[width=11cm]{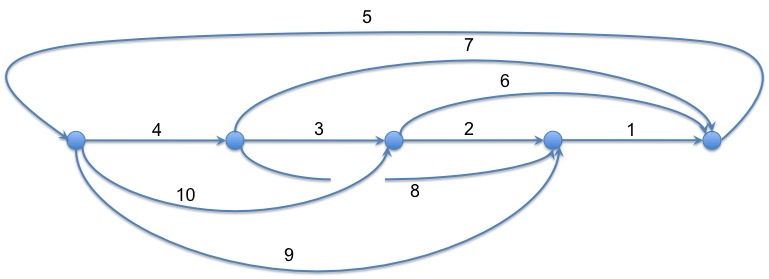}
\caption{The $\G_5^-$ graph.}
\label{gamma5-}
\end{center}
\end{figure}

We also consider the $\G^-_5$ in Fig.\ref{gamma5-} and write down its relativistic q-spin-network
\be
f_{\G_5^-}&:=&\Big[\big[S(\stackrel{\a_5}{\Pi}_{(2)})\big]_{(1)}\stackrel{\a_4}{\Pi}_{(1)}\stackrel{\a_{10}}{\Pi}_{(1)}\stackrel{\a_{9}}{\Pi}_{(1)}\Big]
\otimes \Big[S\big(\stackrel{\a_4}{\Pi}_{(2)}\big)\stackrel{\a_7}{\Pi}_{(1)}\stackrel{\a_3}{\Pi}_{(1)}\stackrel{\a_{8}}{\Pi}_{(1)}\Big]
\otimes\Big[S\big(\stackrel{\a_{10}}{\Pi}_{(3)}\big)S\big(\stackrel{\a_3}{\Pi}_{(2)}\big)\stackrel{\a_6}{\Pi}_{(1)}\stackrel{\a_{2}}{\Pi}_{(1)}\Big]\nonumber\\
&&\otimes\Big[S\big(\stackrel{\a_9}{\Pi}_{(2)}\big)S\big(\stackrel{\a_8}{\Pi}_{(3)}\big)S\big(\stackrel{\a_2}{\Pi}_{(2)}\big)\stackrel{\a_{1}}{\Pi}_{(1)}\Big]
\otimes\Big[S\big(\stackrel{\a_{1}}{\Pi}_{(2)}\big)S\big(\stackrel{\a_6}{\Pi}_{(2)}\big)S\big(\stackrel{\a_7}{\Pi}_{(2)}\big)S\Big(\big[S(\stackrel{\a_5}{\Pi}_{(2)})\big]_{(2)}\Big)\Big]<\stackrel{\a_{5}}{\Pi}_{(1)},\mu^{-1}>
\nonumber\\
&&<\stackrel{\a_{8}}{\Pi}_{(2)}\otimes\stackrel{\a_{10}}{\Pi}_{(2)},\calr^{-1}>
\ee
then by the invariance of Haar integral and the braiding relation,
\be
&&(\id\otimes h^{\otimes4})f_{\G_5^-}\nonumber\\
&=&h\Big[S\big(\stackrel{\a_4}{\Pi}\big)\stackrel{\a_7}{\Pi}_{(1)}\stackrel{\a_3}{\Pi}_{(1)}\stackrel{\a_{8}}{\Pi}_{(1)}\Big]
h\Big[S\big(\stackrel{\a_{10}}{\Pi}_{(2)}\big)S\big(\stackrel{\a_3}{\Pi}_{(2)}\big)\stackrel{\a_6}{\Pi}_{(1)}\stackrel{\a_{2}}{\Pi}_{(1)}\Big]\nonumber\\
&&h\Big[S\big(\stackrel{\a_9}{\Pi}\big)S\big(\stackrel{\a_8}{\Pi}_{(3)}\big)S\big(\stackrel{\a_2}{\Pi}_{(2)}\big)\stackrel{\a_{1}}{\Pi}_{(1)}\Big]
h\Big[S\big(\stackrel{\a_{5}}{\Pi}_{(2)}\big)\ \stackrel{\a_7}{\Pi}_{(2)}\stackrel{\a_6}{\Pi}_{(2)}\stackrel{\a_{1}}{\Pi}_{(2)}\Big]<\stackrel{\a_{5}}{\Pi}_{(1)},\mu^{-1}>
<\stackrel{\a_{8}}{\Pi}_{(2)}\otimes\stackrel{\a_{10}}{\Pi}_{(1)},\calr^{-1}>\label{app3}
\ee
The only difference from $\G^+_5$ relativistic q-spin-network is the previous $\calr$-matrix is replaced by $\calr^{-1}$. Recall that
\be
\langle\ \stackrel{C}{e}_{c}\otimes \stackrel{D}{e}_{d} |\ \stackrel{\alpha}{\Pi}\otimes\stackrel{\b}{\Pi}(\mathcal{R}^{-1})\ | \stackrel{A}{e}_{a}\otimes\stackrel{B}{e}_{b}\ \rangle
&=&\delta_A^C\delta_a^c\sum_{M}\Lambda^{AM}_{DB}(\b)
\left(
                               \begin{array}{cc}
                                 d & c'  \\
                                 D & C \\
                               \end{array}
                             \Bigg|
                             \begin{array}{c}
                                 M  \\
                                 m \\
                               \end{array}
                             \right)
\left(\begin{array}{c}
                                 m   \\
                                 M \\
                               \end{array}
                             \Bigg|\begin{array}{cc}
                                 A & B \\
                                 a' & b \\
                               \end{array}\right)\stackrel{A}{w}_{c'c}(\stackrel{A}{w}\!\!{}^{-1})_{aa'}
\ee
where $\stackrel{A}{w}_{c'c}=\delta_{c',-c}\ q^{-c}(-1)^{A-c'}$ and $(\stackrel{A}{w}\!\!{}^{-1})_{ca'}=\delta_{a',-c} q^c(-1)^{A-a'}$. But the appearance of the factors $\stackrel{A}{w}_{c'c}$ and $(\stackrel{A}{w}\!\!{}^{-1})_{aa'}$ doesn't affect the bound of the $\calr$-matrix representations. Therefore in the same way as the previous arguments for $\G^+_5$ q-spin-network, we obtain:

\begin{Theorem}
The $\G^-_5$ relativistic q-spin-network is integrable, and the vertex amplitude defined by $\G^-_5$ relativistic q-spin-network is finite.
\end{Theorem}

Analogously, we can construct and compute the general vertex amplitudes with general relativistic q-spin-networks. The above arguments can be generalized to general relativistic q-spin-networks, provided that each node of the boundary graph is at least 3-valent. For each 2-valent node there would be a divergent sum $\sum_CC^{n}$, since the $[e^{-2\o C}]^2$ from $|\L^{CB}_{AD}|$ was canceled by the $e^{4\o C}$ from $[d_C]^2$. For a given boundary graph with $N$ nodes, it has ($N-1$) integrals, each of which contributes a $\sum_C (\text{constant})C^{k}e^{-2(n-2)\o C}$. For each crossing, it contributes a $|\sum_M\Lambda^{AM}_{JK}(\a)(2M+1)|$ with is bounded by a polynomial function of $J$ or $K$, while each $J$ or $K$ is linearly bounded by $C$. And note that the factors containing $\mu^{-1}$ doesn't contribute to the infinite sum (in the same way as Eq.\Ref{mu1}) because of Eq.(\ref{mu0}). So the sums over $J$'s and $K$'s at most contribute a polynomial of $C$'s for the bound (each $C$ is associated with a node). Therefore in case that the decaying factor $e^{-2(n-2)\o C}$ doesn't disappear for each node, the sums of $C$'s converges absolutely. As a result,

\begin{Theorem}
The relativistic q-spin-network whose nodes are all at least 3-valent are integrable, and the corresponding vertex amplitude is finite.
\end{Theorem}

\section{A Finite q-Lorentzian Spin-foam Model}

Now we can define a spin-foam model with quantum Lorentz group by the following partition function as a deformation of EPRL spin-foam model:
\be
Z(\ck):=\sum_{K_f=0}^{\frac{\pi}{|\g|\o}}\ \sum_{\nu_e}\ \prod_f\ [d_{K_f}]\ \prod_v\ A_v[{K}_f,\nu_e]
\ee
where $\ck$ is a 2-cell complex, whose vertices all correspond to integrable relativistic q-spin-network graphs. $K_f$ denote the $\Uqsu$ unitary irreps associated to each face and $\nu_e$ denote the $\Uqsu$ intertwiners associated to each edge. One can see immediately that the spin-foam model so defined is \emph{finite}, by the quantum group cut-off $\frac{\pi}{|\g|\o}$ coming from the bound of the $\Uqsl$ unitary irreps.

About the physics from the about finite partition function, we expect that its large-j asymptotics gives a discretized general relativity with positive cosmological constant $\L={\o}/{\ell_p^2}$. More precisely we expect the following result: Given a vertex amplitude $A_v[{K}_f,\nu_e]$ for a 4-simplex (e.g. from $\G_5^+$ or $\G_5^-$ graph), we introduce a parameter $\l$ and replace each $K_f$ by $\l K_f$ and $\o$ by $\o/\l$ ($q=e^{-\o}$). We send $\l\to \infty$ wile keep $K_f\ll\frac{\pi}{|\g|\o}$. In this limit, the vertex amplitude is expected to have the following asymptotic behavior with certain boundary data $({K}_f,\nu_e)$:
\be
A_v[{K}_f,\nu_e]\sim N_1e^{i\l\g S_{\text{Regge},\L}}+N_2e^{-i\l\g S_{\text{Regge},\L}}
\ee
where $N_1,N_2$ are some polynomial functions of $\l$. And $S_{\text{Regge},\L}$ is the Regge action (at the vertex $v$) with a positive cosmological constant $\L={\o}/{\ell_p^2}$ \cite{regge}. But the detailed studies of the asymptotics will be a future research.

\section*{Acknowledgments}

The author thanks C. Rovelli for his encouraging to investigate this topic, and thanks K. Noui and P. Roche for the help about the bound of $\L^{AB}_{CD}$ coefficients. He also would like to thank E. Bianchi, T. Krajewski, T. Masson, and A. Okolow for fruitful discussions.

\section*{Appendix: Haar integral of Relativistic q-spin-networks}

Here we give the detailed derivation for Eq.(\ref{app2}). Given the $\G^+_5$ graph in Fig.\ref{gamma5} and write down its relativistic q-spin-network
\be
f_{\G_5^+}&:=&\Big[\big[S(\stackrel{\a_5}{\Pi}_{(2)})\big]_{(1)}\stackrel{\a_4}{\Pi}_{(1)}\stackrel{\a_{10}}{\Pi}_{(1)}\stackrel{\a_{9}}{\Pi}_{(1)}\Big]
\otimes \Big[S\big(\stackrel{\a_4}{\Pi}_{(2)}\big)\stackrel{\a_7}{\Pi}_{(1)}\stackrel{\a_3}{\Pi}_{(1)}\stackrel{\a_{8}}{\Pi}_{(1)}\Big]
\otimes\Big[S\big(\stackrel{\a_{10}}{\Pi}_{(3)}\big)S\big(\stackrel{\a_3}{\Pi}_{(2)}\big)\stackrel{\a_6}{\Pi}_{(1)}\stackrel{\a_{2}}{\Pi}_{(1)}\Big]\nonumber\\
&&\otimes\Big[S\big(\stackrel{\a_9}{\Pi}_{(2)}\big)S\big(\stackrel{\a_8}{\Pi}_{(3)}\big)S\big(\stackrel{\a_2}{\Pi}_{(2)}\big)\stackrel{\a_{1}}{\Pi}_{(1)}\Big]
\otimes\Big[S\big(\stackrel{\a_{1}}{\Pi}_{(2)}\big)S\big(\stackrel{\a_6}{\Pi}_{(2)}\big)S\big(\stackrel{\a_7}{\Pi}_{(2)}\big)S\Big(\big[S(\stackrel{\a_5}{\Pi}_{(2)})\big]_{(2)}\Big)\Big]<\stackrel{\a_{5}}{\Pi}_{(1)},\mu^{-1}>
\nonumber\\
&&<\stackrel{\a_{10}}{\Pi}_{(2)}\otimes\stackrel{\a_{8}}{\Pi}_{(2)},\calr>
\ee
We are going to show
\be
&&h\Big[S\big(\stackrel{\a_4}{\Pi}\big)\stackrel{\a_7}{\Pi}_{(1)}\stackrel{\a_3}{\Pi}_{(1)}\stackrel{\a_{8}}{\Pi}_{(1)}\Big]
h\Big[S\big(\stackrel{\a_{10}}{\Pi}_{(2)}\big)S\big(\stackrel{\a_3}{\Pi}_{(2)}\big)\stackrel{\a_6}{\Pi}_{(1)}\stackrel{\a_{2}}{\Pi}_{(1)}\Big]\nonumber\\
&&h\Big[S\big(\stackrel{\a_9}{\Pi}\big)S\big(\stackrel{\a_8}{\Pi}_{(3)}\big)S\big(\stackrel{\a_2}{\Pi}_{(2)}\big)\stackrel{\a_{1}}{\Pi}_{(1)}\Big]
h\Big[S\big(\stackrel{\a_{5}}{\Pi}_{(2)}\big)\ \stackrel{\a_7}{\Pi}_{(2)}\stackrel{\a_6}{\Pi}_{(2)}\stackrel{\a_{1}}{\Pi}_{(2)}\Big]<\stackrel{\a_{5}}{\Pi}_{(1)},\mu^{-1}>
<\stackrel{\a_{10}}{\Pi}_{(1)}\otimes\stackrel{\a_{8}}{\Pi}_{(2)},\calr>\nonumber\\
&=&(\id\otimes h^{\otimes4})f_{\G_5^+}
\ee
For each factor, we have from the invariance of Haar integration
\be
&&h\Big[S\big(\stackrel{\a_4}{\Pi}\big)\stackrel{\a_7}{\Pi}_{(1)}\stackrel{\a_3}{\Pi}_{(1)}\stackrel{\a_{8}}{\Pi}_{(1)}\Big]=h\Big[S^{-1}\stackrel{\a_{8}}{\Pi}_{(1)} S^{-1}\stackrel{\a_3}{\Pi}_{(1)} S^{-1}\stackrel{\a_7}{\Pi}_{(1)} \stackrel{\a_4}{\Pi}\Big]\nonumber\\
&=&S^{-1}\stackrel{\a_{8}}{\Pi}_{(2)} S^{-1}\stackrel{\a_3}{\Pi}_{(2)} S^{-1}\stackrel{\a_7}{\Pi}_{(2)} \stackrel{\a_4}{\Pi}_{(1)}h\Big[S\big(\stackrel{\a_4}{\Pi}_{(2)}\big)\stackrel{\a_7}{\Pi}_{(1)}\stackrel{\a_3}{\Pi}_{(1)}\stackrel{\a_{8}}{\Pi}_{(1)}\Big]\label{A1}
\ee
\be
&&h\Big[S\big(\stackrel{\a_{10}}{\Pi}_{(2)}\big)S\big(\stackrel{\a_3}{\Pi}_{(2)}\big)\stackrel{\a_6}{\Pi}_{(1)}\stackrel{\a_{2}}{\Pi}_{(1)}\Big]=h\Big[S^{-1}\stackrel{\a_{2}}{\Pi}_{(1)}S^{-1}\stackrel{\a_6}{\Pi}_{(1)}\stackrel{\a_3}{\Pi}_{(2)}\stackrel{\a_{10}}{\Pi}_{(2)}\Big]\nonumber\\
&=&S^{-1}\stackrel{\a_{2}}{\Pi}_{(2)}S^{-1}\stackrel{\a_6}{\Pi}_{(2)}\stackrel{\a_3}{\Pi}_{(3)}\stackrel{\a_{10}}{\Pi}_{(2)}h\Big[S\big(\stackrel{\a_{10}}{\Pi}_{(3)}\big)S\big(\stackrel{\a_3}{\Pi}_{(2)}\big)\stackrel{\a_6}{\Pi}_{(1)}\stackrel{\a_{2}}{\Pi}_{(1)}\Big]\label{A2}
\ee
\be
&&h\Big[S\big(\stackrel{\a_9}{\Pi}\big)S\big(\stackrel{\a_8}{\Pi}_{(3)}\big)S\big(\stackrel{\a_2}{\Pi}_{(2)}\big)\stackrel{\a_{1}}{\Pi}_{(1)}\Big]=h\Big[S^{-1}\stackrel{\a_{1}}{\Pi}_{(1)}\stackrel{\a_2}{\Pi}_{(2)}\stackrel{\a_8}{\Pi}_{(3)}\stackrel{\a_9}{\Pi}\Big]\nonumber\\
&=&S^{-1}\stackrel{\a_{1}}{\Pi}_{(2)}\stackrel{\a_2}{\Pi}_{(3)}\stackrel{\a_8}{\Pi}_{(4)}\stackrel{\a_9}{\Pi}_{(1)}h\Big[S\big(\stackrel{\a_9}{\Pi}_{(2)}\big)S\big(\stackrel{\a_8}{\Pi}_{(5)}\big)S\big(\stackrel{\a_2}{\Pi}_{(4)}\big)\stackrel{\a_{1}}{\Pi}_{(1)}\Big]\label{A3}
\ee
\be
&&h\Big[S\big(\stackrel{\a_{5}}{\Pi}_{(2)}\big)\ \stackrel{\a_7}{\Pi}_{(2)}\stackrel{\a_6}{\Pi}_{(2)}\stackrel{\a_{1}}{\Pi}_{(2)}\Big]\nonumber\\
&=&S\big(\stackrel{\a_{5}}{\Pi}_{(2)}\big)_{(1)}\ \stackrel{\a_7}{\Pi}_{(3)}\stackrel{\a_6}{\Pi}_{(3)}\stackrel{\a_{1}}{\Pi}_{(3)}h\Big[S\big(\stackrel{\a_{5}}{\Pi}_{(2)}\big)_{(2)}\ \stackrel{\a_7}{\Pi}_{(4)}\stackrel{\a_6}{\Pi}_{(4)}\stackrel{\a_{1}}{\Pi}_{(4)}\Big]\label{A4}
\ee
We only need to check if we neglect the above $h[\cdots]$'s, the rest gives $\big[S(\stackrel{\a_5}{\Pi}_{(2)})\big]_{(1)}\stackrel{\a_4}{\Pi}_{(1)}\stackrel{\a_{10}}{\Pi}_{(1)}\stackrel{\a_{9}}{\Pi}_{(1)}$. We properly insert Eq.(\ref{A3}) into Eq.(\ref{A4})
\be
&&S\big(\stackrel{\a_{5}}{\Pi}_{(2)}\big)_{(1)}\ \stackrel{\a_7}{\Pi}_{(3)}\stackrel{\a_6}{\Pi}_{(3)}\stackrel{\a_{1}}{\Pi}_{(3)}S^{-1}\stackrel{\a_{1}}{\Pi}_{(2)}\stackrel{\a_2}{\Pi}_{(3)}\stackrel{\a_8}{\Pi}_{(4)}\stackrel{\a_9}{\Pi}_{(1)}\nonumber\\
&=&S\big(\stackrel{\a_{5}}{\Pi}_{(2)}\big)_{(1)}\ \stackrel{\a_7}{\Pi}_{(3)}\stackrel{\a_6}{\Pi}_{(3)}\stackrel{\a_2}{\Pi}_{(3)}\stackrel{\a_8}{\Pi}_{(4)}\stackrel{\a_9}{\Pi}_{(1)}
\ee
by $\sum_xx_{(2)}S^{-1}x_{(1)}=\eps(x)1$. Then insert Eq.(\ref{A2}) between $\stackrel{\a_2}{\Pi}_{(3)}$ and $\stackrel{\a_8}{\Pi}_{(4)}$
\be
&&S\big(\stackrel{\a_{5}}{\Pi}_{(2)}\big)_{(1)}\ \stackrel{\a_7}{\Pi}_{(3)}\stackrel{\a_6}{\Pi}_{(3)}\stackrel{\a_2}{\Pi}_{(3)}S^{-1}\stackrel{\a_{2}}{\Pi}_{(2)}S^{-1}\stackrel{\a_6}{\Pi}_{(2)}\stackrel{\a_3}{\Pi}_{(3)}\stackrel{\a_{10}}{\Pi}_{(2)}\stackrel{\a_8}{\Pi}_{(4)}\stackrel{\a_9}{\Pi}_{(1)}\nonumber\\
&=&S\big(\stackrel{\a_{5}}{\Pi}_{(2)}\big)_{(1)}\ \stackrel{\a_7}{\Pi}_{(3)}\stackrel{\a_3}{\Pi}_{(3)}\stackrel{\a_{10}}{\Pi}_{(2)}\stackrel{\a_8}{\Pi}_{(4)}\stackrel{\a_9}{\Pi}_{(1)}
\ee
Now we use the braiding relation
\be
<\stackrel{\a_{10}}{\Pi}_{(1)}\otimes\stackrel{\a_{8}}{\Pi}_{(3)},\calr>\stackrel{\a_{10}}{\Pi}_{(2)}\stackrel{\a_8}{\Pi}_{(4)}=\stackrel{\a_{8}}{\Pi}_{(3)}\stackrel{\a_{10}}{\Pi}_{(1)}<\stackrel{\a_{10}}{\Pi}_{(2)}\otimes\stackrel{\a_{8}}{\Pi}_{(4)},\calr>
\ee
so we have
\be
S\big(\stackrel{\a_{5}}{\Pi}_{(2)}\big)_{(1)}\ \stackrel{\a_7}{\Pi}_{(3)}\stackrel{\a_3}{\Pi}_{(3)}\stackrel{\a_{8}}{\Pi}_{(3)}\stackrel{\a_{10}}{\Pi}_{(1)}\stackrel{\a_9}{\Pi}_{(1)}
\ee
Finally we insert Eq(\ref{A1}) between $\stackrel{\a_{8}}{\Pi}_{(3)}$ and $\stackrel{\a_{10}}{\Pi}_{(1)}$
\be
&&S\big(\stackrel{\a_{5}}{\Pi}_{(2)}\big)_{(1)} \stackrel{\a_7}{\Pi}_{(3)}\stackrel{\a_3}{\Pi}_{(3)}\stackrel{\a_{8}}{\Pi}_{(3)}S^{-1}\stackrel{\a_{8}}{\Pi}_{(2)} S^{-1}\stackrel{\a_3}{\Pi}_{(2)} S^{-1}\stackrel{\a_7}{\Pi}_{(2)} \stackrel{\a_4}{\Pi}_{(1)}\stackrel{\a_{10}}{\Pi}_{(1)}\stackrel{\a_9}{\Pi}_{(1)}\nonumber\\
&=&S\big(\stackrel{\a_{5}}{\Pi}_{(2)}\big)_{(1)} \stackrel{\a_4}{\Pi}_{(1)}\stackrel{\a_{10}}{\Pi}_{(1)}\stackrel{\a_9}{\Pi}_{(1)}
\ee
which proves Eq.(\ref{app2}). For $\G_5^-$ graph, Eq.(\ref{app3}) can be proved in the same way, with a different braiding relation
\be
<\stackrel{\a_{8}}{\Pi}_{(3)}\otimes\stackrel{\a_{10}}{\Pi}_{(1)},\calr^{-1}>\stackrel{\a_{10}}{\Pi}_{(2)}\stackrel{\a_8}{\Pi}_{(4)}=\stackrel{\a_{8}}{\Pi}_{(3)}\stackrel{\a_{10}}{\Pi}_{(1)}<\stackrel{\a_{8}}{\Pi}_{(4)}\otimes\stackrel{\a_{10}}{\Pi}_{(2)},\calr^{-1}>
\ee

\end{document}